\documentclass[11pt,a4paper,english]{article}
\pdfoutput=1 

\usepackage{graphicx}
\usepackage{amssymb,latexsym}
\usepackage{amsmath}
\usepackage{epsf}
\usepackage{pdfsync}
\usepackage{epsfig}
\usepackage[parfill]{parskip}
\usepackage[matrix,arrow,color]{xy}
\usepackage[usenames]{color}
\usepackage[hidelinks]{hyperref} 
\usepackage{bbm}
\usepackage{dsfont}
\usepackage{stmaryrd}
\usepackage{cite}

\usepackage{graphicx} 
\usepackage{pict2e} 

\usepackage{float} 

\usepackage{currvita} 
\usepackage[T1]{fontenc} 

\usepackage{upgreek}

\input{xy}
\xyoption{all}

\input{epsf}

\makeatletter

\catcode`@=11
\renewcommand{\section}
{\@startsection{section}{1}{0pt}{\medskipamount}{\medskipamount}{\large\bf}}
\makeatletter\renewcommand{\subsection}
{\@startsection{subsection}{2}{\z@}{-3.25ex plus -1ex minus -.2ex}
{1.5ex plus .2ex}{\it }}

\numberwithin{equation}{section}
\catcode`@=12

\newcommand{\ba}{\begin{eqnarray*}}
\newcommand{\ea}{\end{eqnarray*}}
\newcommand{\ban}{\begin{eqnarray}}
\newcommand{\ean}{\end{eqnarray}}

\newcommand{\Tr}{\rm{Tr\,}}

\newcommand{\IZ}{\mathbb{Z}}

\newcommand{\cS}{{\cal S}}

\def\e{{\,\rm e}\,}

\def\dd{{\rm d}}

\def\beq{\begin{equation}}
\def\bee{\begin{equation}}
\def\eeq{\end{equation}}
\def\bea{\begin{eqnarray}}
\def\eea{\end{eqnarray}}
\def\bd{\begin{displaymath}}
\def\ed{\end{displaymath}}

\newcommand{\Cint}{\int\kern-10.5pt-\kern7pt}

\newcommand{\be}{\begin{equation}}
\newcommand{\ee}{\end{equation}}

\newcommand\fverbit{\egroup\item[\fbox{\unhbox\pippobox}]}
\newbox\pippobox

\def\w{\wedge}

\def\rm#1{\mathrm{#1}}

\def\be{\begin{equation}}
\def\ee{\end{equation}}
\def\bea{\begin{eqnarray}}
\def\eea{\end{eqnarray}}

\begin{document}

\begin{titlepage}
\setcounter{page}{0}
\begin{flushright}
\small
{\sf RBI-ThPhys-2020-35}
\end{flushright}
\normalsize

\vskip 1.8cm

\begin{center}

{\Large\bf On the R\'enyi entropy of \\[2mm] Lifshitz and hyperscaling violating black holes}

\vspace{15mm}

{\large\bf Zolt\'an K\"ok\'enyesi$^{(a),(b),}$\footnote{ {\tt Zoltan.Kokenyesi@irb.hr}} and Annam{\'a}ria Sinkovics$^{(b),}$\footnote{ {\tt sinkovics@general.elte.hu}}}
\\[6mm]
\noindent{\em $^{(a)}$ Division of Theoretical Physics\\ Ru\dj er Bo\v{s}kovi\'c Institute \\
Bijeni\v{c}ka 54, 10000 Zagreb, Croatia} \\[4mm]
\noindent{\em $^{(b)}$ Institute of Theoretical Physics\\ MTA-ELTE
  Theoretical Research Group \\ E\"otv\"os Lor\'and University \\ P\'azm\'any s. 1/A, 1117
  Budapest, Hungary}

\vspace{20mm}

\begin{abstract}
\noindent
We study R\'enyi entropies for geometries with Lifshitz scaling and hyperscaling violation. We calculate them for specific values of the Lifshitz parameter, and analyze the dual spectrum of the ground state. In the large $d-\theta$ limit they show that the ground state is unique in specific parameter ranges. We also calculate the R\'enyi entropies perturbatively around $n=1$, and derive constraints using the R\'enyi entropy inequalities, which correspond to the thermodynamic stability of the black holes. 
\end{abstract}

\end{center}


\end{titlepage}



{\baselineskip=18pt
\tableofcontents
}

\vspace{5mm}

\section{Introduction}

In the past decades calculating entanglement entropy was extremely useful in quantum information theory, condensed matter physics and quantum chemistry, while it also turned out to be one of the most celebrated aspects of gauge/gravity dualities, which broadened its application to numerous strongly coupled field theories. 

While entanglement entropy (i.e.\ von Neumann entropy) is completely capable to measure entanglement between two subsystems of a pure state, its one parameter deformation, the R\'enyi entropy \cite{Renyi1961,Renyi1965}, carries much more information other than its entanglement characteristics: knowing the R\'enyi entropy for all parameters determines the spectrum of the reduced system. Furthermore, it is much easier to measure experimentally in condensed matter systems \cite{Islam2015}, and it can be used to distinguish between different thermodynamical states, which have the same von Neumann entropy \cite{Lu2017,Dong2018}.

Gravitational dual of R\'enyi entropy in strongly coupled field theories corresponds to introduce conical singularity \cite{Lewkowycz2013,Dong2016} and it was well studied for simple intervals and different thermodynamical ensembles in AdS/CFT \cite{Headrick2010,Myers2011,Fursaev2012,Faulkner2013,Galante2013,Belin2013,Barella2013,Myers2013,Chen2013,Pastras2014,Belin2014,Pastras2015}. The R\'enyi entropy should obey four inequalities by definition \cite{Beck1993,Zyczkowski2003}, which correspond to positivity of thermal entropy and specific heat of a black hole on the dual gravity side \cite{Myers2011,Nakaguchi2016}. These inequalities can give information on less understood dualities, giving constraints on the parameter space of theories that could be connected by holographic duality \cite{Ghodsi2018}.

While the quantum structure of black holes is still not well understood, R\'enyi entropy could be useful to extract information about the black hole microstates at least in the regime where the holographic principle relates them to those of the field theory. Our particular interest is the so called Lifshitz and hyperscaling violating black holes, which are dual to certain condensed matter systems, and our study encompasses their gravitational R\'enyi entropy.

Certain non-relativistic critical systems in condensed matter theory show anisotropic scaling properties between space and time $(t,\vec{x})\rightarrow (\lambda^z t,\lambda \vec{x})$, which is called the Lifshitz scaling, which can be parametrized by the dynamical critical exponent $z$. The corresponding gravity theories realizing this scaling property are given by Lifshitz spacetimes \cite{Kachru2008,Taylor2008,Balasubramanian2009}, which can be thought of as a non-relativistic generalization of the AdS spacetime. One of the importance of this anisotropy is that its specific heat scales at low temperature as $c_V \sim T^{d/z}$, where $d$ is the space dimension of the boundary. Since Fermi liquids show linear dependence such that $c_V \sim T$, Lifshitz scaling theories are good candidate to describe Fermi liquids for $z=d$. 

Theories violating the hyperscaling relations between critical exponents are a one-parameter deformations of the Lifshitz scaling theories and they exhibit a specific heat $c_V \sim T^{(d-\theta)/z}$, where the appearance of the additional hyperscaling violating parameter $\theta$ allows to describe the characteristics of Fermi-liquids for relativistic theories and arbitrary dimensions by using the specific choice $\theta=d-1$ \cite{Gouteraux2011,Ogawa2011,Huijse2011,Alishahiha2012}. Dual spacetimes, which has the corresponding asymptotic scaling property, could be solutions in both Einstein-Proca (see e.g.~\cite{Papadimitriou2014,Papadimitriou2014a,Papadimitriou2014b}) and Einstein-Maxwell-dilaton theories \cite{Taylor2008,Cadani2009,Charmousis2010,Perlmutter2010,Dong2012,Gath2012,Alishahiha2012a,Gouteraux2012,Gouteraux2011}, but the latter has the advantage that it supports analytic black hole solutions for non-zero temperature. Various properties of these solutions and ways to obtain them within supergravity and string theory embeddings were studied in \cite{Alishahiha2012a,Azeyanagi2009,Li2009,Dong2012,Narayan2012,Dey2012,Dey2012a,Perlmutter2012,Cadoni2012,Ammon2012,Kulaxizi2012,Sadeghi2012,Kim2012,Edalati2012,Papadimitriou2018}. These black hole solutions were constructed for planar, spherical and hyperbolic horizon topologies with non-zero charges, and their thermodynamics was extensively surveyed and studied in \cite{Pedraza2018}. Phase transition only occur for spherical topologies with parameter $1\leq z \leq 2$. In grand canonical ensemble, when the electric potential is kept fixed, the phase transition is completely analogous to the Hawking-Page phase transition \cite{Hawking1983,Witten1998}, while in the fixed charge canonical ensemble, it mimics the characteristics of the van der Waals liquid-gas phase transition \cite{Chamblin1999,Chamblin1999a}. 

In Einstein-Maxwell-dilaton realization of hyperscaling violating spacetimes dilaton runs logarithmically, which reflects the fact that these geometries modifies in the deep IR. For flat electrically charged solutions the dilatonic scalar drives the system towards extreme weak coupling in the deep IR, and $\alpha'$ corrections become important. In the presence of magnetic charge the dilaton runs towards strong coupling, and quantum corrections to the gauge kinetic coupling and dilaton potential should be taken into account, which support the emergence of AdS$_2$ in the deep IR \cite{Goldstein2009,Goldstein2010,Sinkovics2012,Harrison2012}. Stability constraint of spatially modulated fluctuations around the IR geometry restricts the form of these quantum corrections \cite{Sinkovics2012a}. Then towards the UV the geometry flows through an intermediate region having hyperscaling violation and Lifshitz scaling, then arrives at AdS$_{d+2}$. 

Black brane geometries with finite temperature have a non-zero horizon, and it is expected that only extremal solutions flow to AdS$_2$ in the deep IR \cite{Alishahiha2012a}. In this paper we study the geometries in the region, where the quantum corrections or other contributions to the gauge kinetic coupling or dilaton potential become relevant. Deeper understanding of their R\'enyi entropy can give further insight on the dual spectrum.

After briefing the necessary background in~\S\ref{sec:BackgroundHypScalBH}, we calculate R\'enyi entropies in~\S\ref{sec:CalcRenyiEntr} for hyperscaling violating black holes in grand canonical ensemble for some integer Lifshitz parameter $z$, and we analyze what insights it can provide on the spectrum of the dual theory. By studying the inequalities we find its stability condition for spherical topology agrees with the one in the case of Hawking-Page phase transition.  We show that the inequalities are not satisfied by geometries with spherical horizon and Lifshitz exponent $1\leq z < 2$ beyond a certain R\'enyi parameter value if the fixed electric potential is smaller than a critical value. We also study R\'enyi entropy in large $d-\theta$ parameter, and discuss the characteristics of the dual ground state in this limit. In~\S\ref{sec:RenyiQCorr} we also calculate R\'enyi entropy perturbatively around $n=1$ (i.e.\ Bekenstein-Hawking entropy), and derive constraints on quantum corrections using the R\'enyi entropy inequalities, which actually correspond to the thermodynamic stability of the black holes.

\vspace{5mm}

\section{Hyperscaling violating black holes}
\label{sec:BackgroundHypScalBH}

In this section we briefly survey the basics on hyperscaling violating and Lifshitz scaling black hole solutions in Einstein-Maxwell-dilaton theory with three Maxwell gauge field. The electric solution and computations of its thermodynamic quantities are borrowed from \cite{Pedraza2018}. We also present its natural generalization to magnetic solution in four dimension, and give a remark on the electric/magnetic duality between the two.

\subsection{Electric solution} 
\label{sec:ElSol}

The action contains kinetic terms for the dilaton field $\phi$ and the three Maxwell fields $A_F$, $A_H$ and $A_K$ with field strengths $F=dA_F$, $H=dA_H$ and $K=dA_K$. They support the solution with Lifshitz scaling ($F$), non-trivial topology ($H$) and non-zero charge ($K$) respectively. The gauge kinetic couplings are given by functions $X(\phi)$, $Y(\phi)$ and $Z(\phi)$ and together with the dilaton potential $V(\phi)$ they determine the action 
\be \label{eq:EMDaction3Maxwell}
\begin{aligned}
\cS_{\rm{EMD}} \, = \, - \frac{1}{16\pi G} \int \dd^{d+2} x \sqrt{-g} & \left(R \, - \, \frac 12 (\partial \phi)^2 \, + \, V(\phi) \, - \, \frac 14 X(\phi) \, F_{\mu\nu}F^{\mu\nu} \right. \\[2mm]
& \left. \qquad \, - \, \frac 14 Y(\phi) \, H_{\mu\nu}H^{\mu\nu} \, - \, \frac 14 Z(\phi) \, K_{\mu\nu}K^{\mu\nu} \right) \ .
\end{aligned}\ee
The equation of motions are 
\be \label{eq:EoMGeneralEMD1} \begin{aligned}
 R_{\mu\nu} \, - \, \frac 12 \, R \, g_{\mu\nu}  \, = &  \, \frac 12 \, \partial_\mu \phi \partial_\nu \phi \, + \, \frac 12 \, g_{\mu\nu} \left(V(\phi) \, - \, \frac 12 \, \partial_\rho \phi \partial^\rho \phi\right) \, - \, \frac 12 \, X(\phi) \left(F_{\mu\rho} {F^\rho}_\nu + \frac 14 \, g_{\mu\nu} F_{\rho\sigma} F^{\rho\sigma} \right) \\[2mm] 
& - \, \frac 12 \, Y(\phi) \left(H_{\mu\rho} {H^\rho}_\nu + \frac 14 \, g_{\mu\nu} H_{\rho\sigma} H^{\rho\sigma} \right) \, - \, \frac 12 \,  Z(\phi) \left(K_{\mu\rho} {K^\rho}_\nu + \frac 14 \, g_{\mu\nu} K_{\rho\sigma} K^{\rho\sigma} \right) \ ,
\end{aligned} \ee
\be \label{eq:EoMGeneralEMD2}
D_\mu \partial^\mu \phi \, + \, \partial_\phi V(\phi) \, - \, \frac 14 \, \partial_\phi X(\phi) F_{\mu\nu} F^{\mu \nu} \, - \, \frac 14 \, \partial_\phi Y(\phi) H_{\mu\nu} H^{\mu \nu} \, - \, \frac 14 \, \partial_\phi Z(\phi) K_{\mu\nu} K^{\mu \nu} \, = \, 0 \ ,
\ee \vspace{-0.5mm}
\be \label{eq:EoMGeneralEMD3}
D_\mu \left(X(\phi) \, F^{\mu \nu} \right) = 0 \ , \qquad D_\mu \left(Y(\phi) \, H^{\mu \nu} \right) = 0  \qquad  \text{and} \qquad D_\mu \left(Z(\phi) \, K^{\mu \nu} \right) = 0 \ .
\ee 
Here we focus on electric solutions, we come back to the magnetic case in $d=2$ later. The hyperscaling violating and Lifshitz scaling solution is given by the black hole metric
\be \label{eq:HypeLifmetricGenDim}
d s^2 \, = \, \left( \frac{r}{r_F}\right)^{-\frac{2\theta}{d}} 	\left(-\left(\frac{r}{\ell}\right)^{2z} f(r) \, d t^2 \, + \, \frac{\ell^2}{f(r)r^2} \, d r^2 \, + \, r^2 \, d \Omega_{k,d}^2\right) \ ,
\ee
with blackening factor
\be \label{eq:BlackeningFactorGenDimZTh}
f(r) \, = \, 1 - \frac{m}{r^{d-\theta+z}} + \frac{q^2}{r^{2(d-\theta+z-1)}} + k\,\frac{(d-1)^2}{(d-\theta+z-2)^2} \, \frac{\ell^2}{r^2} \ .
\ee
The black hole parameters introduced here are the mass parameter $m$ and charge parameter $q$, and $\ell$ is the overall scale of the geometry.  $r_F$ is the upper cut-off, but it does not play any further role in this paper. 
The horizon part of the metric $d \Omega_{k,d}^2$ is defined differently for planar ($k=0$), spherical ($k=1$) and hyperbolical ($k=-1$) topologies, such that
\be
\begin{aligned}
d \Omega^2_{0,d} \, &= \, \frac{d x_0^2}{\ell^2} + \ldots + \frac{d x_{d-1}^2}{\ell^2} \ , \\[2mm]
d \Omega^2_{1,d} \, &= \, d x_0^2 + \sin^2 \! x_0 \, d x_1^2 + \ldots + \sin^2\! x_0 \dots \sin^2\! x_{d-2} \, d x_{d-1}^2 \ , \\[2mm]
d \Omega^2_{-1,d} \, &= \, d x_0^2 + \sinh^2\! x_0 \, d \Omega^2_{1,d-1}  \ . 
\end{aligned}
\ee
Although the same notion of horizon coordinates was used here for different topologies, they do not range the same. For planar topology they are usual compact variables, while for spherical and hyperbolical horizons they are the respective standard angles, and they both have a regularized finite volume denoted by $\omega_{k,d} := \int d\Omega_{k,d}^2$. The geometry is considered to be valid only in an intermediate region between the IR (close to the horizon) and the UV (close to the asymptotic boundary at $r\rightarrow \infty$, the cut-off scale is given by $r_F$). We require that the blackening factor $f(r)\rightarrow 1$ as $r$ approaches the boundary, which is satisfied by using the constraint $d-\theta+z > 0$ for neutral solution, and $d-\theta+z - 1 > 0$,  if the charge parameter is finite. The metric \eqref{eq:HypeLifmetricGenDim} in this limit has a scaling symmetry
\be
t  \, \rightarrow \, \lambda^z \, t \ , \qquad \Omega \, \rightarrow \, \lambda \, \Omega \ , \qquad r \, \rightarrow \, \lambda^{-1} \, r  \qquad \text{and} \qquad ds \, \rightarrow \, \lambda^{\theta/d} \, ds \ . 
\ee   

The electric field strengths are given in terms of functions $E_F(r)$, $E_H(r)$ and $E_K(r)$ with 
\be
F = E_F(r) \, dt \w dr \ , \qquad H = E_H(r) \, dt \w dr \qquad \text{and} \qquad K=E_K(r) dt \w dr \ ,
\ee 
and their dependence on the radial coordinate can be derived from the equation of motions, such as 
\be 
E_F(r) = E_{F,0} \, r^{d-\theta+z-1} \ , \qquad E_H(r) = E_{H,0} \, r^{d-\theta+z-3}  \qquad \text{and} \qquad E_K(r) = E_{K,0} \, r^{-(d-\theta+z-1)} \ .
\ee
The solution of the dilaton has the logarithmic form
\be \label{eq:DilatonLogGamma}
\phi = \phi_0 + \gamma \log r \qquad \text{with} \qquad \gamma=\sqrt{2(d-\theta)(z-1-\theta/d)} \ .
\ee 
The gauge kinetic functions and dilaton potential, which support the metric above, is a first order exponential such that
\be \label{eq:GaugeKinDilPotAnsatz}
X(\phi) = X_0 \e^{\! 2\alpha_X \phi} \ , \qquad  Y(\phi) = Y_0 \e^{\! 2\alpha_Y \phi}  \ , \qquad Z(\phi) = Z_0 \e^{\! 2\alpha_Z \phi} \qquad \text{and} \qquad V(\phi) = V_0 \e^{\! \eta \phi} \ .
\ee 
and the coefficients are given by
\be \label{eq:GaugeKinDilPotExpression} \begin{aligned}
X_0 &= \frac{2(z-1)(d-\theta+z)}{\ell^{2z} \, E_{F,0}^2}\, r_F^{2\theta/d} \e^{\! -2\alpha_X \phi_0} \ , \qquad &\alpha_X &= - \frac{(d-\theta+\theta/d)}{\gamma} \ , \\[2mm]
Y_0 &= \frac{2k \, (d-1)(d(z-1)-\theta)}{(d-\theta+z-2)\,\ell^{2(z-1)} \, E_{H,0}^2}\, r_F^{2\theta/d} \e^{-2\alpha_Y \phi_0} \ , \qquad &\alpha_Y &= - \frac{(d-1)(d-\theta)}{d\gamma} \ , \\[2mm]
Z_0 &= \frac{2q^2 \, (d-\theta)(d-\theta+z-2)}{\ell^{2z} \, E_{K,0}^2}\, r_F^{2\theta/d} \e^{\! -2\alpha_Z \phi_0} \ , \qquad &\alpha_Z &= \frac{z-1-\theta/d}{\gamma} \ , \\[2mm]
V_0 &= \frac{(d-\theta+z-1)(d-\theta+z)}{\ell^{2} \, r_F^{2\theta/d}	}\, \e^{\! -\eta \phi_0} \ , \qquad &\eta &= \frac{2\theta}{d\gamma  } \ . \\
\end{aligned}\ee
 
\subsection{Magnetic solution in $d=2$ and electric/magnetic duality}
\label{sec:MagneticElMagnDual}

The magnetic solution is only known in four dimensions. We take the magnetic field strengths to be constants with respect to the radial coordinate, and we define them by
\be
F \, = \, Q_F \, \varphi_k(x_0) \, dx_0 \w dx_1 \, , \qquad  H \, = \, Q_H \, \varphi_k(x_0) \, dx_0 \w dx_1 \qquad \text{and} \qquad K \, = \, Q_K \, \varphi_k(x_0) \, dx_0 \w dx_1 \, ,  
\ee
where the function $\varphi_k(x_0)$ distinguishes the different topologies such that
\be
\varphi(x_0) \, = \, \left\{ \begin{aligned}
\ & 1, \quad & \text{if}  & \quad k=0 \ , \\[2mm]
\ & \ell^2  \sin x_0 , \quad & \text{if} & \quad k=1 \ , \\[2mm]
\ & \ell^2  \sinh x_0 , \quad & \text{if} & \quad k=-1 \ .
\end{aligned} \right. 
\ee
The magnetic solution is slightly differs from the electric. The metric, dilaton and dilaton potential are the same as those were given for the electric solution in $d=2$ by \eqref{eq:HypeLifmetricGenDim}, \eqref{eq:BlackeningFactorGenDimZTh}, \eqref{eq:DilatonLogGamma}, \eqref{eq:GaugeKinDilPotAnsatz} and \eqref{eq:GaugeKinDilPotExpression}, while the gauge kinetic functions describe an inverse coupling to the field strengths as
\be 
X(\phi) = X_0 \e^{\! - 2\alpha_X \phi} \ , \qquad  Y(\phi) = Y_0 \e^{\! - 2\alpha_Y \phi}  \qquad \text{and} \qquad Z(\phi) = Z_0 \e^{\! - 2\alpha_Z \phi} \ .
\ee 
The coupling is inverse in the sense that a gauge field that was weekly coupled in the electric case is now strongly coupled and vica versa. The coefficients are slightly different than the electric ones, and they are given by 
\be \label{eq:GaugeKinExpressionMagn}  \begin{aligned}
X_0 &= \frac{2(z-1)(2-\theta+z)}{\ell^{6} \, Q_{F}^2}\, r_F^{\theta} \e^{\! 2\alpha_X \phi_0} \ ,\\[2mm]
Y_0 &= \frac{2k \, (2(z-1)-\theta)}{(z-\theta)\,\ell^{4} \, Q_{H}^2}\, r_F^{\theta} \e^{2\alpha_Y \phi_0} \ , \\[2mm]
Z_0 &= \frac{2q^2 \, (2-\theta)(z-\theta)}{\ell^{6} \, Q_{K}^2}\, r_F^{\theta} \e^{\! 2\alpha_Z \phi_0} \ , \\
\end{aligned}\ee
and the exponential constants $\alpha_X$, $\alpha_Y$ and $\alpha_Z$ are the same as those previously defined by \eqref{eq:GaugeKinDilPotExpression}.

{\underline{\sl Electric/magnetic duality in $d=2$.} \ }
One can see that both the electric and magnetic field strengths support the same geometry, but they are not dual to each other in general. Following \cite{Hawking1995}the electric/magnetic duality $F \leftrightarrow \e^{2\alpha \phi} *\! F$ with $\phi \leftrightarrow - \phi$ is only true if the dilaton potential $V(\phi)$ is a constant. If we denote the electric field strength and coupling by $F^{(e)}=E_F(r) \, dt\w dr$ and $X^{(e)}(\phi)$, the magnetic ones by $F^{(m)}=Q_F \, dx \w dy$ and $X^{(m)}(\phi)$, the duality can be formulated precisely as
\be \label{eq:ElMagnChangeSol2d}
F^{(m)} \, = \, \e^{2 \alpha_X \phi} * F^{(e)} \ , \qquad F^{(e)} \, = \, -  \e^{-2 \alpha_X \phi} * F^{(m)} \qquad \text{and} \qquad X^{(m)}(\phi) \, = \, X^{(e)}(-\phi) \ ,
\ee
and analogously for the other two gauge fields. The duality changes the corresponding gauge term in the action by a sign such that
\be
X^{(e)}(\phi) \, F^{(e)}_{\mu\nu} F^{(e)\, \mu\nu} \, = \, - X^{(m)}(\phi) \, F^{(m)}_{\mu\nu} F^{(m)\, \mu\nu} \ .
\ee
The expressions for the field strengths \eqref{eq:GaugeKinDilPotExpression} and \eqref{eq:GaugeKinExpressionMagn} yield the correspondence 
\be
Q_F \, = \, - \ell^{z-3}\e^{2\alpha_X \phi_0} E_{F,0} \ , \qquad Q_H \, = \, - \ell^{z-3}\e^{2\alpha_Y \phi_0} E_{H,0} \qquad \text{and} \qquad Q_F \, = \, - \ell^{z-3}\e^{2\alpha_Z \phi_0} E_{K,0} 
\ee 
between magnetic and electric constants. If the dilaton potential is not constant, the change between the electric and magnetic solutions defined in \eqref{eq:ElMagnChangeSol2d} can not be derived by using a field redefinition $\phi\leftrightarrow -\phi$. The constant dilaton potential yields the vanishing of the hyperscaling violating coefficient $\theta$, hence pure Lifshitz scaling geometries in four dimensions have electric/magnetic duality. 

Another possibility is that if $V(\phi)\equiv 0$, which is satisfied for $\theta=z+1$ or $\theta=z+2$. This would spoil the asymptotic scaling of the geometry, but can be consistent with a UV completion, if the quantum corrections for $V(\phi)$ support AdS$_4$ in the UV.

\subsection{Null energy condition}

The null energy condition is required by the duality in order to have a reasonable field theory on the boundary. It is a constraint on the energy-momentum tensor, which says $ T_{\mu\nu} n^\mu n^\nu \geq 0$ for arbitrary null-vector $n_\mu$. The energy-momentum tensor is given by the Einstein tensor with $T_{\mu\nu}=R_{\mu\nu} - \frac 12 g_{\mu\nu}R$. By choosing two orthogonal null-vectors the condition gives two inequalities
\be
\begin{aligned}
0 \, & \leq \, (d-\theta)(d(z-1)-\theta)  \ , \\[2mm]
0 \, & \leq \,  \frac{r^2}{\ell^2}\, (z-1)(d-\theta+z) \, + \, k \, \frac{(d-1)(d(z-1)-\theta)}{(d-\theta+z-2)} \, + \, q^2 \, \frac{(d-\theta)(d(z-1)-\theta)}{\ell^2 r^{2(d-\theta+z-2)}} \ ,
\end{aligned}
\ee
which should hold for arbitrary radius.

Since we expect that the energy scale of the dual field theory ranges between the horizon, which is located at the radius $r_h$, and the UV cut-off, which we take to be at infinite radius here, we require the null energy condition to be satisfied on this domain. We discuss two further limits.

{\underline{\sl Finite horizon radius in the deep IR.} \ }
By requiring that $f(r)$ asymptotes to 1 gave the constraint $d-\theta+z > 0$.  At large radius the second inequality yields $z\geq 1$, while it gives a more involved expression for finite $r_h$. In this case the null-energy conditions are summarized as
\be\label{eq:NECfiniteRad}\begin{aligned}
1 \, & \leq \, z \ , \\[2mm]
0 \, & \leq \, (d-\theta)(d(z-1)-\theta)  \ , \\[2mm]
0 \, & \leq \,  \frac{r_h^2}{\ell^2}\, (z-1)(d-\theta+z) \, + \, k \, \frac{(d-1)(d(z-1)-\theta)}{(d-\theta+z-2)} \, + \, q^2 \, \frac{(d-\theta)(d(z-1)-\theta)}{\ell^2 r_h^{2(d-\theta+z-2)}} \qquad \text{for $k\neq 0$} \ ,
\end{aligned} \ee 

{\underline{\sl Horizon radius goes to zero in the deep IR.} \ } The limit when the second inequality is considered to be hold between $r\rightarrow 0 $ and $r \rightarrow \infty$ was discussed in \cite{Pedraza2018}. They assumed $d-\theta+z-2 > 0$ and $d-\theta > 0$, then arrived at the null-energy conditions
\be \label{eq:NECPedraza}
z \, \geq \, 1 \ , \quad \qquad d(z-1)-\theta \, \geq \, 0 \quad \qquad \text{and} \quad \qquad k(d(z-1)-\theta) \, \geq \, 0 \ .
\ee
The third inequality gives the hyperscaling violating exponent a fix value $\theta=d(z-1)$ for hyperbolic horizons, while it is not relevant for the other two cases. This means the factor $\gamma$ goes to zero for hyperbolic topologies, and thus the fields need to be rescaled in order to have a reasonable solution. On the level of the action the limit $\gamma\rightarrow 0$ yields a zero kinetic term, a constant gauge kinetic funtion $Z(\phi)$ and the vanishing of the gauge field $H$. 

\subsection{Thermodynamics}

In the following we briefly describe the thermodynamics of the above introduced black hole solutions. Here we only focus on the case when the electric potential is fixed on the boundary, which is also called the grand canonical ensemble. The reader can find more detailed information together with the description of the fixed charge ensemble (canonical ensemble) in \cite{Pedraza2018}. Since the magnetic potential does not appear in the thermodynamic first law, there is no difference between the two ensembles in that case. We also mention for clarity that the thermodynamic potentials for electric and magnetic solutions agree in the canonical ensemble.

We express the black hole temperature in $d$ dimensions by using the horizon radius, which is defined as the largest root of $f(r_h)=0$, giving 
\be \label{eq:MassParameterHorizon}
m \, = \, r_h^{d-\theta+z} \left(1 \, + \, \frac{q^2}{r_h^{2(d-\theta+z-1)}} \, + \, k\, \frac{(d-1)^2}{(d-\theta+z-2)^2} \frac{\ell^2}{r_h^2} \right) \ .
\ee 
The temperature can be calculated by using the standard Eucklidean trick, which gives 
\be \label{eq:TempLHBH}
\begin{aligned}
T \, &= \, \frac{|f'(r_h)|}{4\pi} \left(\frac{r_h}{\ell}\right)^{z+1} \\[2mm] 
& = \, \frac{r_h^z}{4\pi \ell^{z+1}} \left( (d-\theta+z) \, - \, q^2 \, \frac{(d-\theta+z+2)}{r_h^{2(d-\theta+z-1)}} \, + \, k\, \frac{(d-1)^2}{(d-\theta+z-2)} \frac{\ell^2}{r_h^2} \right) \ ,
\end{aligned}\ee
where we assumed $f'(r_h) \geq 0$, otherwise a minus sign should appear in the expression above. The thermal entropy is given by the Bekenstein-Hawking formula
\be \label{eq:BHEntropy}
S = \frac{\omega_{k,d} \, r_F^\theta}{4G} \, r_h^{d-\theta} \ .
\ee
The mass of the black hole, which appears in the first law, is computed by the ADM mass formula \cite{Arnowitt1960,Arnowitt1962} on the asymptotic boundary after proper renormalization. The ADM mass on a fixed $r$ radial slice of a constant time surface is given by
\be \label{eq:ADMmassDef}
M_{\rm{ADM}} \, = \, - \, \frac{1}{8 \pi G} \int_{S_{k,d}} d^d x \frac{\sqrt{-g_{tt}}}{\sqrt{g_{rr}}} \partial_r \sqrt{\sigma} \ .
\ee
where $\sigma$ is the determinant of the induced metric on $S_{k,d}$ that is a radial slice at $r=R$ of a constant time surface. The actual mass is calculated after the renormalization, which depends on the ground state of the ensemble.

The field strengths $F$ and $H$ support the asymptotic scaling and topology of the internal space, hence the corresponding charges need to be kept fixed, otherwise the boundary theory would be ill-defined. Thus only the charge corresponding to $K$ can be varied on the boundary in grand canonical ensemble, which is 
\be \label{eq:BHcharge}
Q \, = \, \frac{1}{16 \pi G} \int Z(\phi) *K \, = \, \frac{\omega_{k,d}}{16 \pi G} \e^{\!\alpha_Z \phi_0} \sqrt{2Z_0(d-\theta)(d-\theta+z-2)} \, q \, \ell^{-1} r_F^{\theta-\theta/d} \ .
\ee
The electric gauge one-form $A_K$ is chosen that way it vanishes on the horizon, which is satisfied by
\be
A_K \, = \, \frac{E_{K,0}}{d-\theta+z-2} \left( \frac{1}{r_h^{d-\theta+z-2}} - \frac{1}{r^{d-\theta+z-2}}\right) dt \ .
\ee
Then the electric potential $\Phi$ is computed as the asymptotic value of the gauge field $A_K$, which is
\be \label{eq:ElPot}
\Phi \, = \, \frac{E_{K,0}}{d-\theta+z-2} \, \frac{1}{r_h^{d-\theta+z-2}} \ .
\ee
Here we used the condition $d-\theta+z-2 > 0$ in order to have a well-defined electric potential on the boundary.

{\underline{\sl Electric solution in grand canonical ensemble.} \ } 
The ground state is the extremal black hole (i.e.~vanishing temperature) with zero charge. For planar and spherical topologies ($k=0,1$) this corresponds to vanishing horizon ($r_h=0$), but in the case of hyperbolic solution $k=-1$ the horizon is not zero, and it induces a negative mass parameter. We use the following notation for both cases
\be \begin{aligned}
r_{h,\text{ground}} \, & = \,
\left\{ \ \begin{aligned}
 & \, 0 \qquad &\text{for $k=0,1$} \ , \\[2mm]
  &  \sqrt{\frac{\ell^2(d-1)}{(2-z)(z+d(2-z))}} \qquad &\text{for $k=-1$} \ , \end{aligned} \right. \\[2mm]
m_{\text{ground}} \, & = \,
\left\{ \ \begin{aligned}
 & \, 0 \qquad &\text{for $k=0,1$} \ , \\[2mm]
  & - \frac{2\ell^2 \, r_{h,\text{ground}}^{(d-1)(2-z)}}{(2-z)^2(z+d(2-z))} \qquad &\text{for $k=-1$} \ \end{aligned} \right.
\end{aligned}\ee
The ADM mass is computed by using the background subtraction method as a renormalization scheme. The background is the ground state heated up to a temperature $T^{(0)}$, which is subtracted from the excited state (black hole with parameters $m$, $q$, $r_h$ and $T$), while both of them are calculated with a given radial cutoff $R$. The temperatures $T$ and $T^{(0)}$ are matched to each other in order to have the same reach at $R$ regarding the Euclidean time direction, giving the expression for large $R$
\be
\frac{1}{T^{(0)}} = \frac{1}{T} \left(1- \frac{m-m_{\text{ground}}}{2R^{d-\theta+z}} \right) \ .
\ee
Then the ADM mass \eqref{eq:ADMmassDef} gives
\be \label{eq:ADMGrandCan}
M \, = \, \frac{\omega_{k,d}}{16\pi G} \, \frac{r_F^{\theta}(d-\theta)}{\ell^{z+1}} \,  (m-m_{\text{ground}}) \ .
\ee
Since the potential is kept fixed, the temperature as well as other thermodynamic quantities that depend on the charge parameter $q$ should be expressed in terms of electric potential $\Phi$ given in \eqref{eq:ElPot}. Then the temperature \eqref{eq:TempLHBH} reads as
\be  \label{eq:TempBHPhi}
T \, = \, \frac{r_h^z}{4\pi \ell^{z+1}} \left( (d-\theta+z) \, + \, (d-\theta + z - 2) \left( \Phi_c^2 - \Phi^2\right) \frac{c^2}{r_h^2} \right) \ ,
\ee
where the constant $c=(d-\theta+z-2)q/E_{K,0}$, and we used the notation
\be \label{eq:PhicDef}
\Phi_c^2 \, = \, k \, \frac{(d-1)^2}{(d-\theta+z-2)^2} \frac{\ell^2}{c^2} \ ,
\ee
which also denotes a critical value of the electric potential and it have a role in phase transition for $k=1$, which will become clear later. The charge \eqref{eq:BHcharge} in terms of the electric potential is the following
\be
Q \, = \, \frac{\omega_{k,d}}{16\pi G} \frac{r_F^\theta}{\ell^{z+1}} \, 2(d-\theta)(d-\theta+z-2) \, c^2 \Phi \, r_h^{d-\theta+z-2} \ .
\ee

The thermodynamical potential in this ensemble is the Gibbs potential
\be \label{eq:GibbsPotDef}
G \, = \, M  -  TS - \Phi Q\ , 
\ee
where $M$ is computed by the ADM Mass \eqref{eq:ADMGrandCan}, $T$ and $S$ are the black hole temperature and entropy given in \eqref{eq:TempLHBH} and \eqref{eq:BHEntropy}, while the potential $\Phi$ and charge $Q$ are derived in \eqref{eq:ElPot} and \eqref{eq:BHcharge}. Finally the Gibbs potential yields
\be \label{eq:GibbsPot}
G \, = \, \frac{\omega_{k,d}}{16\pi G} \, \frac{r_F^{\theta}}{\ell^{z+1}}  \left[ r_h^{d-\theta + z} \left(-z + (2-z)\left(\Phi_c^2 - \Phi^2\right)\frac{c^2}{r_h^2}\right) \, - \, (d- \theta) \, m_{\text{ground}} \right] \ .
\ee
The thermodynamic potential defined in this way agrees with the renormalized Euclidean on-shell action of \eqref{eq:EMDaction3Maxwell} with the corresponding boundary terms (see \cite{Pedraza2018}) and divided by the temperature.

{\underline{\sl Magnetic solution in canonical ensemble.} \ } 
The ground state corresponds to the extremal black hole with mass parameter
\be
m_{\text{ext}} \, = \, 2 r_{\text{ext}}^{2-\theta+z}\left(\frac{1+z-\theta}{z-\theta} + \frac{k}{(z-\theta)^2}\frac{\ell^2}{r_{\text{ext}}^2}\right) \ ,
\ee
where the extremal horizon satisfies $f(r_{\text{ext}})=f'(r_{\text{ext}})=0$. In order to calculate the thermodynamic mass, the extremal background with temperature determined by
\be
\frac{1}{T^{(0)}} = \frac{1}{T} \left(1- \frac{m-m_{\text{ext}}}{2R^{d-\theta+z}} \right)
\ee
at large $R$ is subtracted from the ADM mass \eqref{eq:ADMmassDef}, and it gives
\be
M \, = \, \frac{\omega_{k,d}}{16\pi G} \, \frac{r_F^{\theta}(2-\theta)}{\ell^{z+1}} \,  (m-m_{\text{ext}}) \ .
\ee
The thermodynamic potential in canonical ensemble is the Helmholtz free energy defined by
\be
F \, = \, M  -  TS \ .
\ee
The temperature follows from \eqref{eq:TempLHBH} in $d=2$, and $S$ is given by the Bekenstein-Hawking entropy in \eqref{eq:BHEntropy}. Thus the free energy reduces to
\be \begin{aligned}
F \, &=& \, - \, \frac{\omega_{k,2}}{16\pi G} \frac{r_F^\theta}{\ell^{z+1}} \, \frac{1}{T} & \bigg( z\big(r_h^{2-\theta+z} - r_{\text{ext}}^{2-\theta+z}\big) - k \ell^2 \frac{(2-z)}{(z-\theta)^2}\big(r_h^{z-\theta} - r_{\text{ext}}^{z-\theta} \big)  \\[2mm]
& & & \quad - q^2 (2(1-\theta) + z) \big(r_h^{-(z-\theta)} - r_{\text{ext}}^{-(z-\theta)} \big) \bigg) \\[3mm]
     &=& \, - \, \frac{\omega_{k,2}}{16\pi G} \frac{r_F^\theta}{\ell^{z+1}} \, \frac{1}{T} &   \bigg( z r_h^{2-\theta+z} - q^2 (2(1-\theta) + z) r_h^{-(z-\theta)} -k \ell^2 \frac{(z-2)}{(z-\theta)^2} \, r_h^{z-\theta} + (2-\theta) m_{\text{ext}}  \bigg) \ .
\end{aligned}\ee
We note here that the free energy in canonical ensemble agrees for electric and magnetic fluxes, since both the thermodynamic variables and ground state are considered to be the same.

\vspace{5mm}

\section{Calculation of R\'enyi entropy}
\label{sec:CalcRenyiEntr}

In this sections we review the holographic calculation and inequalities of R\'enyi entropy mostly based on \cite{Myers2011}, and calculate it for special values of the Lifshitz exponent. We also check the inequalities for general Lifshitz and hyperscaling violating parameters and relate them to the thermodynamic stability and phase transitions known in the literature (see e.g. \cite{Pedraza2018}).

\subsection{Holographic R\'enyi entropy}

An arbitrary quantum state can be written as a thermal state 
\be \label{eq:normThermalstate}
\rho = \frac{\e^{- H_{\rm{mod}}/T}}{\Tr\!\e^{-H_{\rm{mod}}/T}}
\ee
by introducing a modular Hamiltonian $H_{\rm{mod}}$. Here we study states dual to black hole solutions we described in the previous section. The different thermodynamical ensembles correspond to different modular Hamiltonians such that
\be \begin{aligned}
H_{\rm{mod}} \, &= \, H & \qquad &\text{for canonical ensemble,} \\[2mm]
H_{\rm{mod}} \, &= \, H - \Phi Q & \qquad &\text{for grand canonical ensemble,}
\end{aligned}\ee
where $H$ and $Q$ here are understood as the physical Hamiltonian and conserved charge operators in the dual theory, and $\Phi$ is the electric potential. The temperature of the thermal state agrees with the Hawking temperature. The partition functions $\Tr\!\e^{-H_{\rm{mod}}/T}$ give the corresponding thermodynamic potentials by reducing to $\e^{\! - F/T}$ or $\e^{\! - G/T}$ respectively. The von Neumann entropy of a state $\rho$ is given by
\be
S \, = \, - \, \Tr\rho\log\rho \ ,
\ee
and the R\'enyi entropy is its deformation by an extra parameter $n$, and it is defined as
\be \label{eq:RenyiDef}
S_n = \frac{1}{1-n} \log \Tr \rho^n \ .
\ee
Some specific values of $n$ captures relevant information about the dual theory. In the $n\rightarrow 1$ limit it reduces to the von Neumann entropy, which in the context of black holes should agree with the Bekenstein-Hawking entropy. The $n\rightarrow 0$ limit it gives the logarithm of the number of non-vanishing eigenvalues or the rank of density operator in the case of a discrete spectrum, which is expected to be divergent. The third limit, which could be relevant, is $n\rightarrow \infty$. Then the R\'enyi entropy gives $-\log \lambda_1$, where $\lambda_1$ is the largest eigenvalue of the density operator. It also calculates the ground state energy $E_1$ of the modular Hamiltonian by $S_\infty=(E_1-F)/T$ or $(E_1-G)/T$ depending on the thermodynamic ensemble. In general it can have multiple degeneracies, which is specified by an integer number if the spectrum is discrete or by a spectral density in the case of continuous spectrum. It is possible and also expected that the spectrum has both discrete and continuous parts (see e.g.~\cite{Myers2011} or~\S\ref{sec:ExDiscContSpec}). After expanding the R\'enyi entropy for discrete spectrum around $n\rightarrow\infty$, one arrives at 
\be
S_n \, = \, - \log \lambda_1  -  \frac{1}{n}\log(d(\lambda_1)\lambda_1)  + \mathcal{O}\left(\frac{1}{n^2},\frac{1}{n}\left(\frac{\lambda_2}{\lambda_1}\right)^n\right) \ , 
\ee
where the degeneracy of $\lambda_1$ is denoted by $d(\lambda_1)$. So one can see that the $1/n$ term in the expansion is related to the degeneracy in a way that
\be
\log d(\lambda_1) = \big( S_n + n^2 \partial_n S_n \big)\big|_{n=\infty} \ ,
\ee  
if the expression we have for $S_n$ is analytic. Another way to calculate the degeneracy of $\lambda_1$ follows from the entropy of the ground state such that $S_{\rm{gr}}=\log d(\lambda_1)$. The expansion described above works correspondingly for continuous spectrum.
 
The R\'enyi entropy of a reasonable quantum theory should be positive and satisfy the following four inequalities
\be\begin{aligned} \label{eq:RenyiIneq}
\frac{\partial S_n}{\partial n} \, & \leq  \,  0 \ , \\[2mm]
\frac{\partial }{\partial n}\left(\frac{n-1}{n} \, S_n\right) \, & \geq \,  0 \ , \\[2mm]
\frac{\partial }{\partial n}\big((n-1) \, S_n\big) \, & \geq \,  0 \ , \\[2mm]
\frac{\partial^2 }{\partial n^2}\big((n-1) \, S_n\big) \, & \leq \,  0 \ . \\[2mm]
\end{aligned}\ee
The second and third inequalities are coming from the positivity of entropy, while the first and fourth hold as long as the system has a positive specific heat. 

One can calculate the R\'enyi entropy of field theories by introducing an $n$-sheeted branched cover of the geometry, which circles around the original space $n$ times and branches over the entanglement surface. On the dual side this branched cover is computed by a regular bulk geometry, which asymptotes to the branched cover on the boundary. To analitically continue away from integer values of $n$, one introduces an orbifold geometry by factorizing the regular bulk manifold with the replica symmetry $\IZ_n$, which cyclically permutes the bulk sheets. This construction introduces a conical singularity with deficit angle $2\pi (1-1/n)$, which can be continue away from integer numbers (see \cite{Dong2016}). Then the R\'enyi entropy can be expressed as
\be \label{eq:RenyiEntrFromAction}
S_n = \frac{1}{n-1} \, \big( I(n) - n I(1) \big) \ ,
\ee
where $I(n)$ in the classical limit  is given by the renormalized on-shell Euclidean bulk action of the regular covering geometry. Practically $I(n)$ is computed by the redefinition of Euclidean time period as $\tau \sim \tau + n/T$ and the corresponding horizon $r_h(n)$ resulting in a regular geometry. This means in the present study that $r_h(n):=r_h(T/n)$, if the horizon $r_h(T)$ is understood as a function of the temperature on a connected domain including both $T$ and $T/n$. 

By calculating the on-shell Euclidean bulk action results in the thermodynamic potential of the corresponding ensemble divided by the temperature (see e.g. \cite{Pedraza2018}). Hence the R\'enyi entropy for canonical and grand canonical ensembles are given by
\be \label{eq:RenyiEntThermoPot} \begin{aligned}
S_n \, &= \, \frac{n}{n-1} \frac{1}{T} \big(F(T/n) - F(T) \big) \qquad &&\text{canonical ensemble,} \\[2mm]
S_n \, &= \, \frac{n}{n-1} \frac{1}{T} \big(G(T/n) - G(T) \big) \qquad &&\text{grand canonical ensemble.}
\end{aligned}\ee 
Following \cite{Myers2011} one can rewrite these expressions using the thermodynamical formulas of thermal entropy 
\be \begin{aligned}
S  &=  - \left(\frac{\partial F}{\partial T}\right)_Q  \qquad &&\text{canonical ensemble,} \\[2mm]
 S  &= - \left(\frac{\partial G}{\partial T}\right)_\Phi \qquad &&\text{grand canonical ensemble.}
\end{aligned} \ee
One arrives at
\be \label{eq:RenyiEntrExpressionInt}
S_n \, = \, \frac{n}{n-1} \frac{1}{T} \int_{T/n}^T S(T') dT' \ ,
\ee
where the integration is understood as that the respective thermodynamical variable is kept fixed. One can see that the R\'enyi entropy contains information about all of the thermal entropies with temperature ranging between $T$ and $T/n$. While the positivity of $S(T/n)$ for all R\'enyi parameter values between $n$ and 1 ensures that $S_n$ is positive as expected, conversely this is not true in general. The condition $S_n \geq 0$ for a fixed value of $T$ can be satisfied by a system with its thermal entropy having negative values while its integrand is positive. This feature can be resolved by the R\'enyi entropy inequalities \eqref{eq:RenyiIneq}. The second inequality precisely gives the constraint $S(T/n) \geq 0$. The situation with the first and fourth inequality is somewhat similar. The fourth one gives the constraint that the specific heat
\be
C_{Q/\Phi}(T) = T \left(\frac{\partial S}{\partial T}\right)_{Q/\Phi}
\ee
at $T/n$ is positive for all $n$ under consideration. It also results in the first inequality, but the equivalent condition to the first one is the positivity of the integrand $\int_{T/n}^T(S(T') - S(T/n)) dT'$, which could enable negative specific heat for some $n$. The third inequality is satisfied by using the first and the second ones. So if the R\'enyi entropy has the form \eqref{eq:RenyiEntrExpressionInt}, the inequalities and its positivity at a given temperature can be rewritten as two conditions 
\be \label{eq:condEntrSpecHeatPos}
S\big(T/n\big) \geq 0 \qquad \text{and} \qquad C_{Q/\Phi}\big(T/n\big) \geq 0   \ ,
\ee
which should be satisfied for all $n$ under consideration.

In practice we calculate \eqref{eq:RenyiEntrExpressionInt} as
\be \label{eq:RenyiIntegralRh}
S_n \, = \, \frac{n}{n-1} \frac{1}{T} \int_{r_h(n)}^{r_h(1)} S(r_h) \frac{\partial T}{\partial r_h} dr_h \ ,
\ee
which translates the conditions \eqref{eq:condEntrSpecHeatPos} into
\be \label{eq:condEntrSpecHeatPosRh}
S\big(r_h(n)\big) \geq 0 \qquad \text{and} \qquad \bigg[\left(\frac{\partial S}{\partial r_h}\right)_{Q/\Phi} \left(\frac{\partial T}{\partial r_h}\right)^{-1}\bigg] \bigg|_{r_h(n)} \geq 0 \ .
\ee

\subsection{R\'enyi entropy of electric solution in grand canonical ensemble}

In the following we present our calculations on the R\'enyi entropy for hyperscaling violating and Lifshitz scaling black hole geometries discussed in the previous section. We show that R\'enyi entropy inequalities are not satisfied by geometries with spherical horizon and Lifshitz exponent $1\leq z < 2$ beyond a certain R\'enyi parameter value if the fixed electric potential is smaller than a critical value.

The R\'enyi entropy corresponding to the geometry discussed previously in~\S\ref{sec:ElSol} can be calculated by using the formulas either \eqref{eq:RenyiEntThermoPot} or \eqref{eq:RenyiIntegralRh}. To use the first formula one needs the Gibbs potential computed in \eqref{eq:GibbsPot}, while the second one can be derived from the black hole temperature \eqref{eq:TempBHPhi} and entropy \eqref{eq:BHEntropy}. They both give the result 
\be
S_n \, = \, \frac{n}{n-1}\frac{\omega_{k,d}}{16\pi G} \, \frac{r_F^{\theta}}{\ell^{\theta-d+1}} \, \frac{1}{T}  \left[ z\big(x^{d-\theta + z} - x_n^{d-\theta + z} \big) + \frac{c^2}{\ell^2}(2-z)\left(\Phi_c^2 - \Phi^2\right)\left(x_n^{d-\theta+z-2}-x^{d-\theta+z-2}\right) \right] ,
\ee
where we introduced the notation $x_n:=r_h(n)/\ell$ and $x:=x_1$ for simplicity. So the quantity $x_n$ is the horizon solution to the $n$-sheeted bulk geometry with temperature $T/n$ and normalized by the length scale $\ell$. In general it is difficult to compute $x_n$ analytically, therefore we will only study it qualitatively for arbitrary parameters, and quantitatively for specific parameter choices.

One can see that the R\'enyi entropy is positive by rewriting the expression as an integral over $n$ and using the property 
\be \label{eq:ConstrxndTdx}
\left. x_n'\, \frac{\partial T}{\partial x}\right|_{x_n} = - T/n^2 \leq 0 \ ,
\ee
where $x_n'=dx_n/dn$. 

The non-trivial constraint, which comes from the inequalities is the positivity of specific heat. Since $\partial S/\partial r_h \geq 0$, it follows from \eqref{eq:condEntrSpecHeatPosRh} and \eqref{eq:ConstrxndTdx} that 
\be
\left. \frac{\partial T}{\partial x}\right|_{x_n} \, \geq \, 0 \ \qquad \Leftrightarrow \qquad x_n' \leq 0 \ ,
\ee
which means the temperature of the black hole increases with the horizon radius, and $x_n$ decreases with the R\'enyi parameter.

By analyzing the possible horizon solutions of the equation
\be \label{eq:TperneqTxn}
\frac{T}{n} \, = \, \mathcal{T}(x_n) \ ,
\ee
where the function $\mathcal{T}(x)$ is defined by the temperature as a function of horizon radius given in \eqref{eq:TempBHPhi}, one can see that it has different number of roots $x_n$ with respect to the value of potential $\Phi$, horizon topology $k$ and Lifshitz exponent $z$.  We discuss them in four separate cases. In the following we use the null-energy conditions given in \eqref{eq:NECPedraza}, which assume that the horizon is sufficiently small in the IR.

{\underline{\sl The case $1 \leq z < 2$, and $\Phi_c^2 < \Phi^2$.} \ }  It includes solutions with flat and hyperbolic horizon topologies $(k=0,-1)$ and spherical topology $(k=1)$ with electric potential $\Phi>\Phi_c$. The derivative $d \mathcal{T} / d x_n$ is always positive, which means all solutions are stable and $\mathcal{T}(x_n)$ is strictly increasing function. Its minimal value for finite $T$ is given by 
\be \label{eq:xinfty}
x_\infty \, = \, \frac{|c|}{\ell}\sqrt{\frac{d-\theta+z-2}{d-\theta+z}(\Phi^2-\Phi_c^2)} \ .
\ee
Thus there is one and only one solution for all $n>0$, so the R\'enyi entropy satisfies all the inequalities within this parameter range. The horizon of the $n$-sheeted bulk geometry goes to a finite value, so the R\'enyi entropy approaches a finite value as $n$ increases. 

{\underline{\sl The case $2 < z$ and $\Phi_c^2 < \Phi^2$.} \ } The horizon topologies that satisfy the condition are the same ($k=0,-1$ and $k=1$ with $\Phi>\Phi_c$). The temperature function $\mathcal{T}(x_n)$ vanishes at the $x_\infty$ given in \eqref{eq:xinfty}. The derivative $d \mathcal{T} / d x_n$ changes sign where $\mathcal{T}(x_n)$ takes negative value, so all solutions larger than $x_\infty$ are stable, thus there are exactly one allowed horizon $x_n$ for all $n>0$. So the R\'eny entropy in this parameter range is qualitatively identical to the previous one, it exists and not restricted by the inequalities for all value of R\'enyi parameter, while goes to a finite value for large $n$. 

{\underline{\sl The case $1 \leq z < 2$ and $\Phi^2 < \Phi_c^2$.} \ }  The only possible horizon topology is the spherical $k=1$, while the electric potential $\Phi$ is smaller than the critical value $\Phi_c$. The function $\mathcal{T}(x_n)$ is always positive, and it has a minimal value at the point
\be 
x_{n_{\text{max}}} \, = \, \frac{|c|}{\ell} \sqrt{\frac{(2-z)}{z}\frac{d-\theta+z-2}{d-\theta+z}(\Phi^2_c - \Phi^2)} \ ,
\ee  
where its derivative changes sign. Roots $x_n<x_{n_{\text{max}}}$ are not stable, as they give $d \mathcal{T} / d x_n < 0$, hence they are not allowed by the R\'enyi entropy inequalities. The maximal value of R\'enyi parameter corresponding to $x_{n_{\text{max}}}$ is
\be \label{eq:nmaxBH}
n_{\text{max}} \,  = \, 2 \pi \ell^{z+1} T \, \frac{2-z}{d-\theta+z} \left(\frac{(2-z)}{z}\frac{d-\theta+z-2}{d-\theta+z} \, c^2(\Phi^2_c - \Phi^2)\right)^{-z/2} \ ,
\ee
so the the allowed value of the parameter ranges from $0$ to $n_{\text{max}}$. For higher parameters the inequalities are no longer satisfied.

{\underline{\sl The case $2 < z$ and $\Phi^2 < \Phi_c^2$.} \ } Again the geometry has to be supported by spherical horizon topology. The derivative $d \mathcal{T}/ d x_n$ is always positive, and $\mathcal{T}(x_n)$ vanishes at zero horizon $x_\infty=0$, so all values of R\'enyi parameter are allowed by the inequalities.
The R\'enyi entropy behaves qualitatively the same as the first two parameter cases, it goes to a finite value, although this time the horizon of the $n$-sheeted bulk geometry approaches zero for large $n$.

We have seen that geometries with spherical horizon topology and Lifshitz exponent $1 \leq z < 2$ support qualitatively different holographic R\'enyi entropies than the ones on other parameter ranges, if the fixed electric potential is smaller than a critical value. The inequalities limit the R\'enyi parameter in the holographic calculation, which leads to a maximal allowed value $n_{\text{max}}$. This is directly related to a thermodynamic instability of the geometry, which occurs below a finite horizon and leads to a minimal value of the temperature (see e.g.~\cite{Pedraza2018}). In our discussion the R\'enyi parameter have a similar role as the inverse temperature in the context of thermodynamic stability. So the appearance of an upper bound of R\'enyi parameter indicates an instability of the $n$-sheeted bulk geometry used in holographic calculation.

\subsection{Calculation of R\'enyi entropy for specific values of the Lifshitz parameter}

In this subsection we present our analytic computations of R\'enyi entropy for the first few integer values of the Lifshitz parameter. The results could be applied for example to study the background geometries of holographic superconducting fluctuations \cite{Sin2009,Brynjolfsson2009} or other holographic condensed matter systems. 

In order to calculate the R\'enyi entropy, we need the roots of the algebraic equation \eqref{eq:TperneqTxn}, which can be computed analytically for a few specific values of Lifshitz parameter $z$. It reduces to second order equation for $z=1,2,4$, third order for $z=3,6$ and forth order for $z=3/2,8$. In the following we focus on the first four integer values $z=1,2,3,4$.

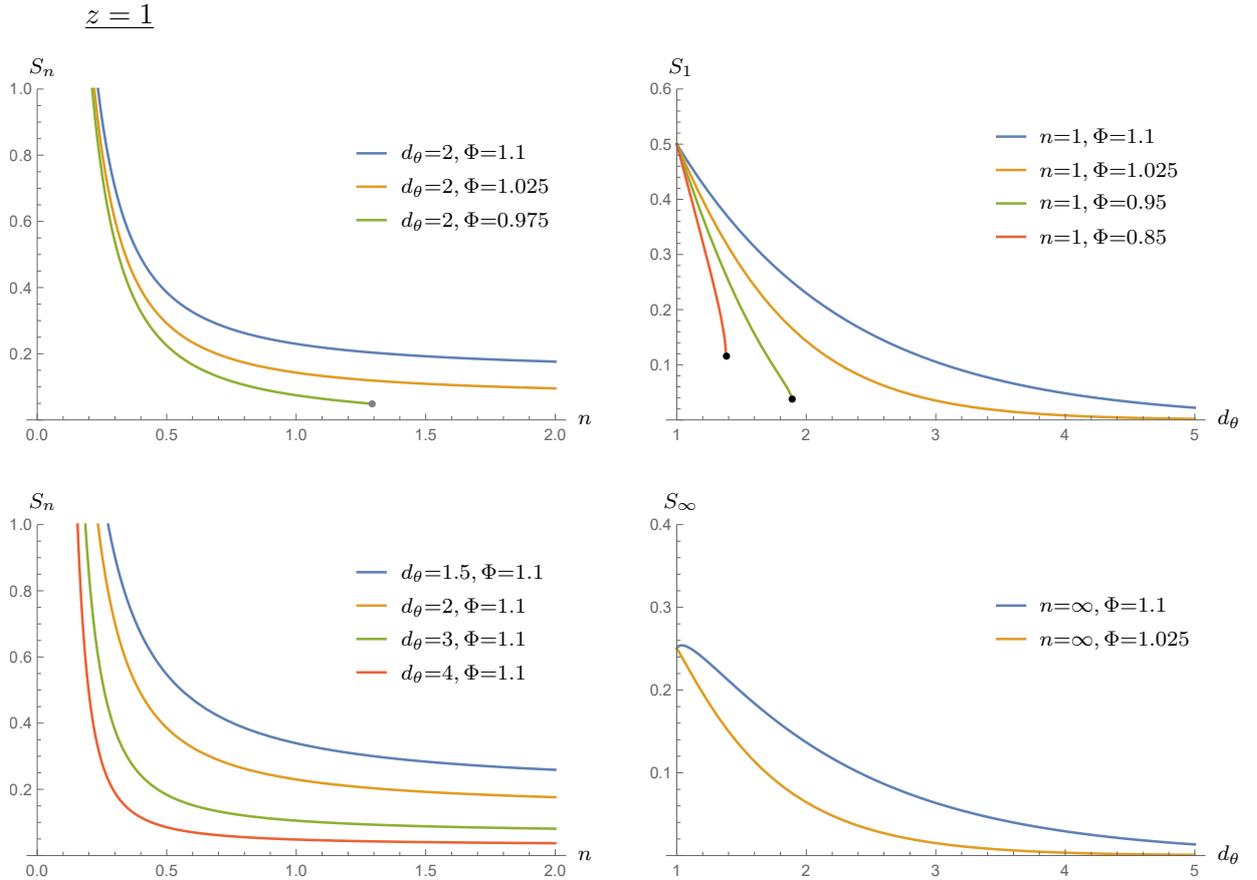
\begin{figure}[htb]
   \centering
  \setlength{\unitlength}{0.1\textwidth}
   \begin{picture}(10,7.2)
     \put(0.1,0){\includegraphics[width=0.46\textwidth]{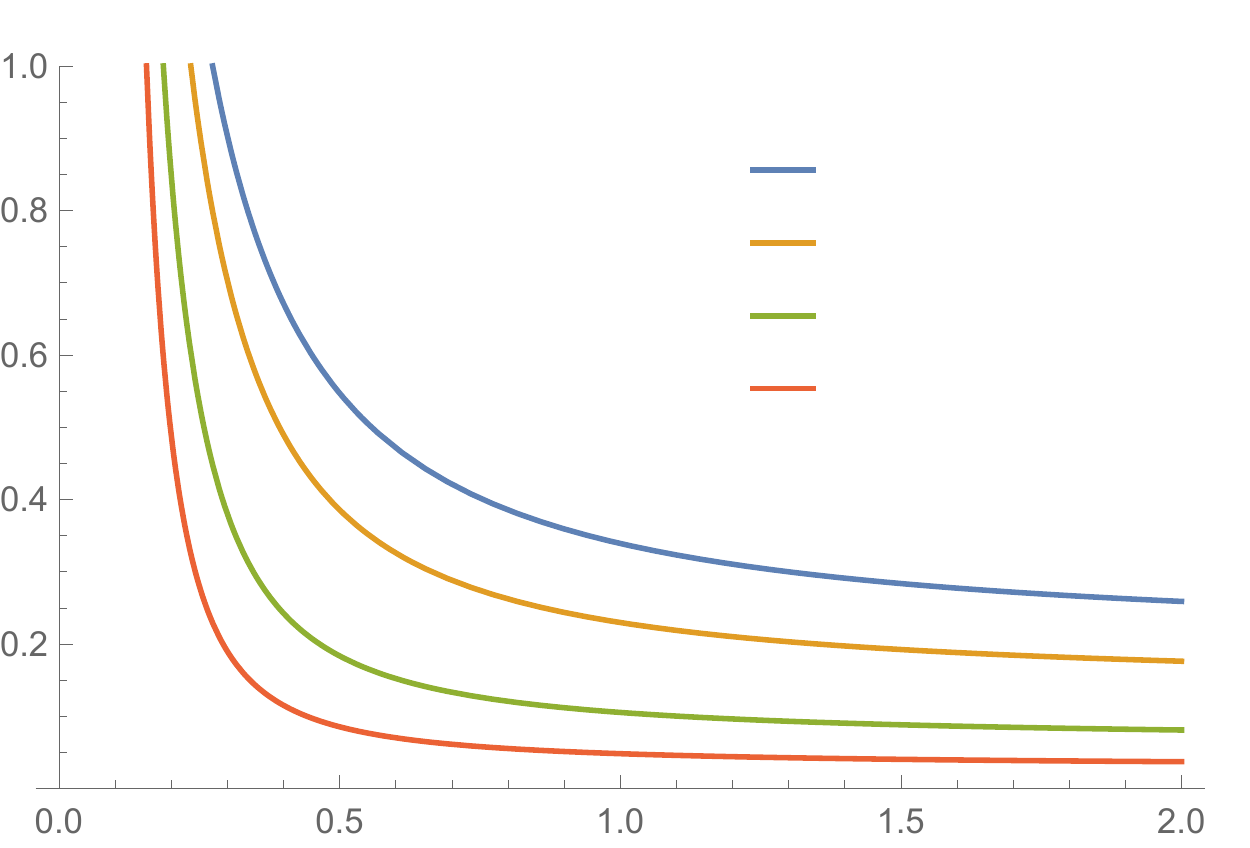}}
		\put(0.1,3.5){\includegraphics[width=0.46\textwidth]{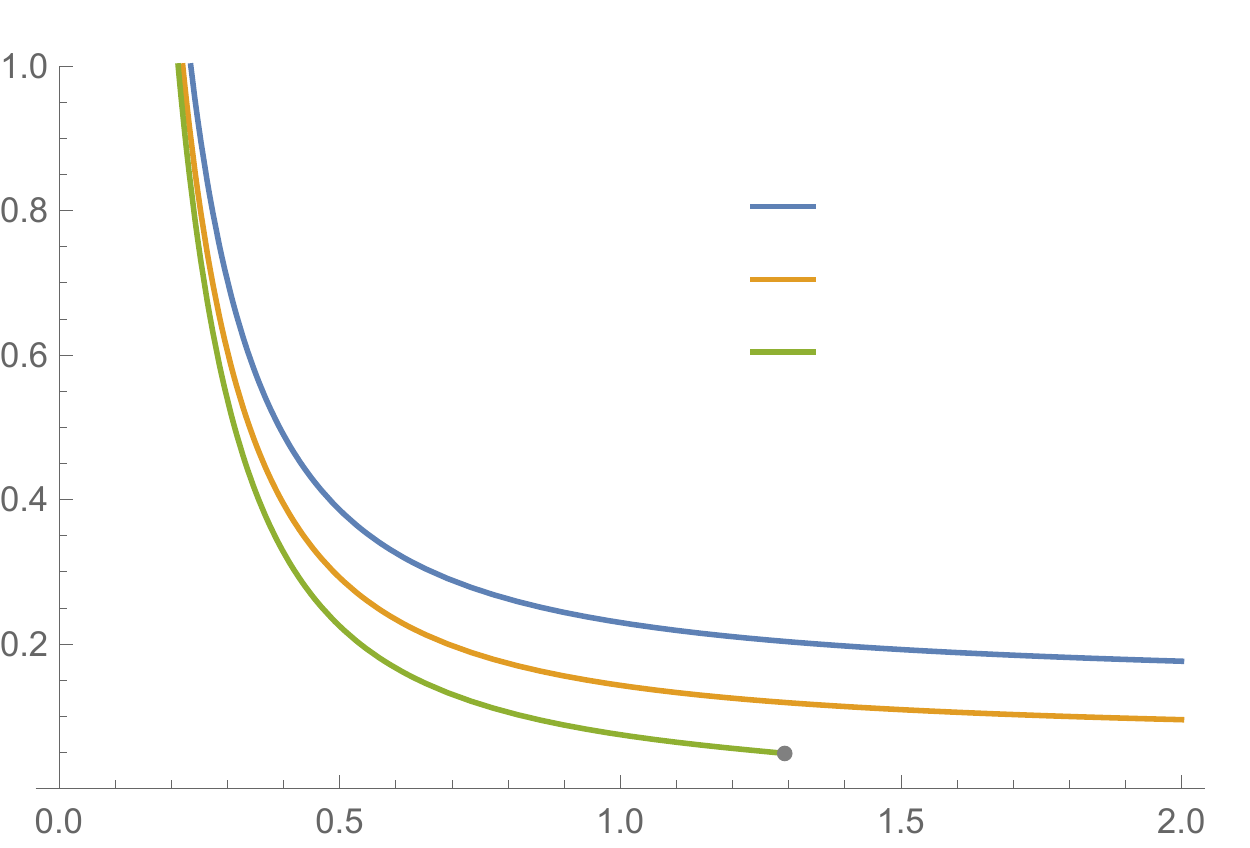}}
		\put(5.2,0){\includegraphics[width=0.46\textwidth]{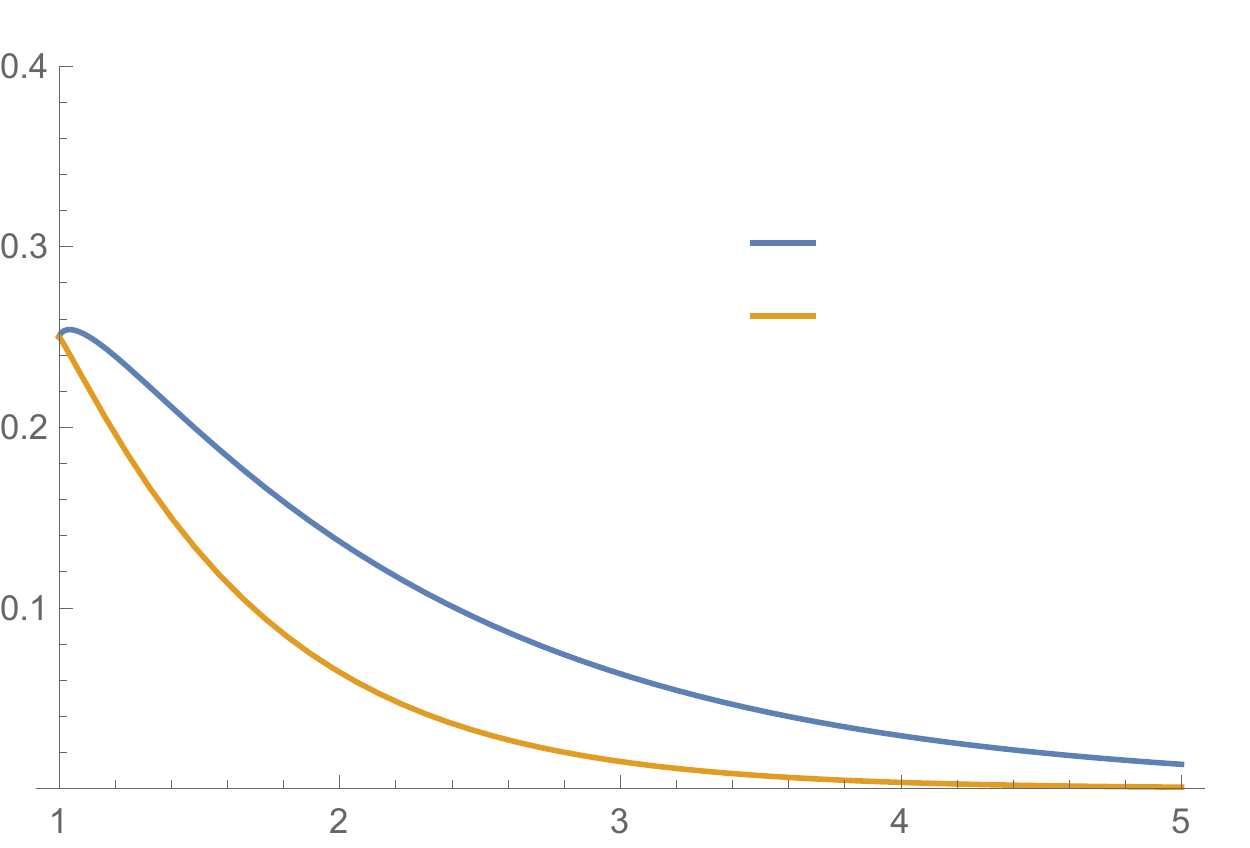}}
		\put(5.2,3.5){\includegraphics[width=0.46\textwidth]{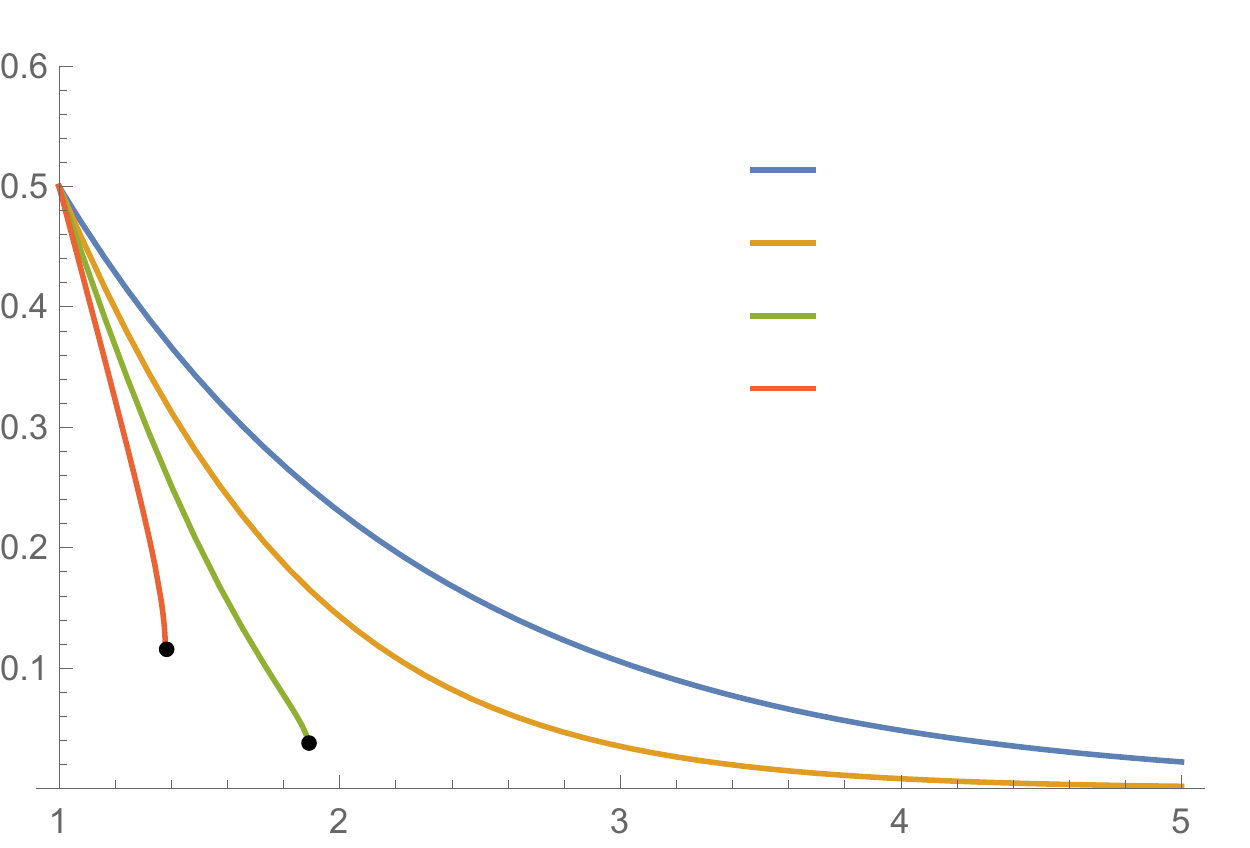}}
     \put(0.7,6.9){\underline{$z=1$}}
		\put(0.25,6.5){${\scriptstyle S_n}$}
		\put(4.63,3.68){${\scriptstyle n}$}
		\put(3.22,5.81){${\scriptstyle d_\theta=2, \, \Phi=1.1}$}
		\put(3.22,5.54){${\scriptstyle d_\theta=2, \, \Phi=1.025}$}
		\put(3.22,5.265){${\scriptstyle d_\theta=2, \, \Phi=0.975}$}
		\put(0.25,3.01){${\scriptstyle S_n}$}
		\put(4.63,0.18){${\scriptstyle n}$}
		\put(3.22,2.442){${\scriptstyle d_\theta=1.5, \, \Phi=1.1}$}
		\put(3.22,2.168){${\scriptstyle d_\theta=2, \, \Phi=1.1}$}
		\put(3.22,1.902){${\scriptstyle d_\theta=3, \, \Phi=1.1}$}
		\put(3.22,1.632){${\scriptstyle d_\theta=4, \, \Phi=1.1}$}
		\put(5.35,6.5){${\scriptstyle S_1}$}
		\put(9.73,3.68){${\scriptstyle d_\theta}$}
		\put(8.31,5.942){${\scriptstyle n=1, \, \Phi=1.1}$}
		\put(8.31,5.672){${\scriptstyle n=1, \, \Phi=1.025}$}
		\put(8.31,5.410){${\scriptstyle n=1, \, \Phi=0.95}$}
		\put(8.31,5.133){${\scriptstyle n=1, \, \Phi=0.85}$}
		\put(5.31,3.01){${\scriptstyle S_\infty}$}
		\put(9.73,0.18){${\scriptstyle d_\theta}$}
		\put(8.31,2.177){${\scriptstyle n=\infty, \, \Phi=1.1}$}
		\put(8.31,1.903){${\scriptstyle n=\infty, \, \Phi=1.025}$}
   \end{picture}
   \caption{R\'enyi entropies are presented with parameter choices $z=1$, $T=1/4\pi$, $\omega_{k,d}=4G$ and $r_F,\ell,c,\Phi_c^2=1$. The first column shows its dependence on the R\'enyi parameter with fix values of $d_\theta=d-\theta$ and $\Phi$. The upper left panel illustrates the case, when there is a maximal value of the parameter $n$, which is represented as a gray dot. In the second column the plots show the $n\rightarrow 1$ and $n\rightarrow \infty$ limits vs.~$d_\theta=d-\theta$. The black dots on the upper right panel indicate maximal values for $d_\theta$. Higher values would lead to non-real horizons. One can see if $\Phi^2>\Phi^2_c$ then both $S_1$ and $S_\infty$ goes to zero. The latter suggests that the largest eigenvalue $\lambda_1$ of the density matrix goes towards one at large $d_\theta$.}
   \label{fig:Plotz1}
\end{figure}

{\underline{\sl The case $z=1$.} \ }
Close to the asymptotic boundary this geometry has a relativistic scaling, and it simplifies to the AdS geometry if the hyperscaling violating parameter $\theta$ is zero. This case is the same as was studied in \cite{Myers2013}. The root $x_n$ of the algebraic equation of the temperature \eqref{eq:TperneqTxn} for all values of $\Phi$ is given by 
\be
x_n \, = \, \frac{\frac{4\pi\ell T}{n}+\sqrt{\left(\frac{4\pi\ell T}{n}\right)^2-4(d-\theta+1)(d-\theta-1)c^2\ell^{-2}(\Phi_c^2-\Phi^2)}}{2(d-\theta+1)} \ .
\ee
If the expression under the square root is negative, the solution does not exist, which is only possible if the horizon has spherical topology and $\Phi<\Phi_c$. It gives the maximal value for the R\'enyi parameter 
\be 
n_{\text{max}} \,  = \, \frac{2 \pi \ell^2 T}{\sqrt{(d-\theta+1)(d-\theta-1)c^2(\Phi_c^2-\Phi^2)}} \ ,
\ee

which agrees with \eqref{eq:nmaxBH}. This is just the third case we discussed in the previous subsection: the $n$-sheeted bulk geometry is thermodynamically instable for $n>n_{\text{max}}$.

In Figure~\ref{fig:Plotz1} we illustrate how the R\'enyi entropy decreases with $n$ at some specific values of electric potential and the shifted dimension $d_\theta = d-\theta$ by the hyperscaling violating parameter. The plot shows that the large $n$ limit $S_\infty$ decreases with $\Phi$ until it reaches the value corresponding to the critical potential $\Phi_c$ (for spherical topology). We also show the qualitative behavior of $S_1$ and $S_\infty$ with respect to $d_\theta$. The Bekenstein-Hawking entropy $S_1$ is important as it must be positive, and $S_\infty$ gives $-\log\lambda_1$, where $\lambda_1$ is the largest eigenvalue of dual density matrix $\rho$. In other words $S_\infty T$ is the smallest eigenvalue or ground state of the dual modular Hamiltonian. The plot illustrates how the ground state energy of the modular Hamiltonian approaches zero.

\begin{figure}[htb]
   \centering
  \setlength{\unitlength}{0.1\textwidth}
   \begin{picture}(10,7.2)
     \put(0.1,0){\includegraphics[width=0.46\textwidth]{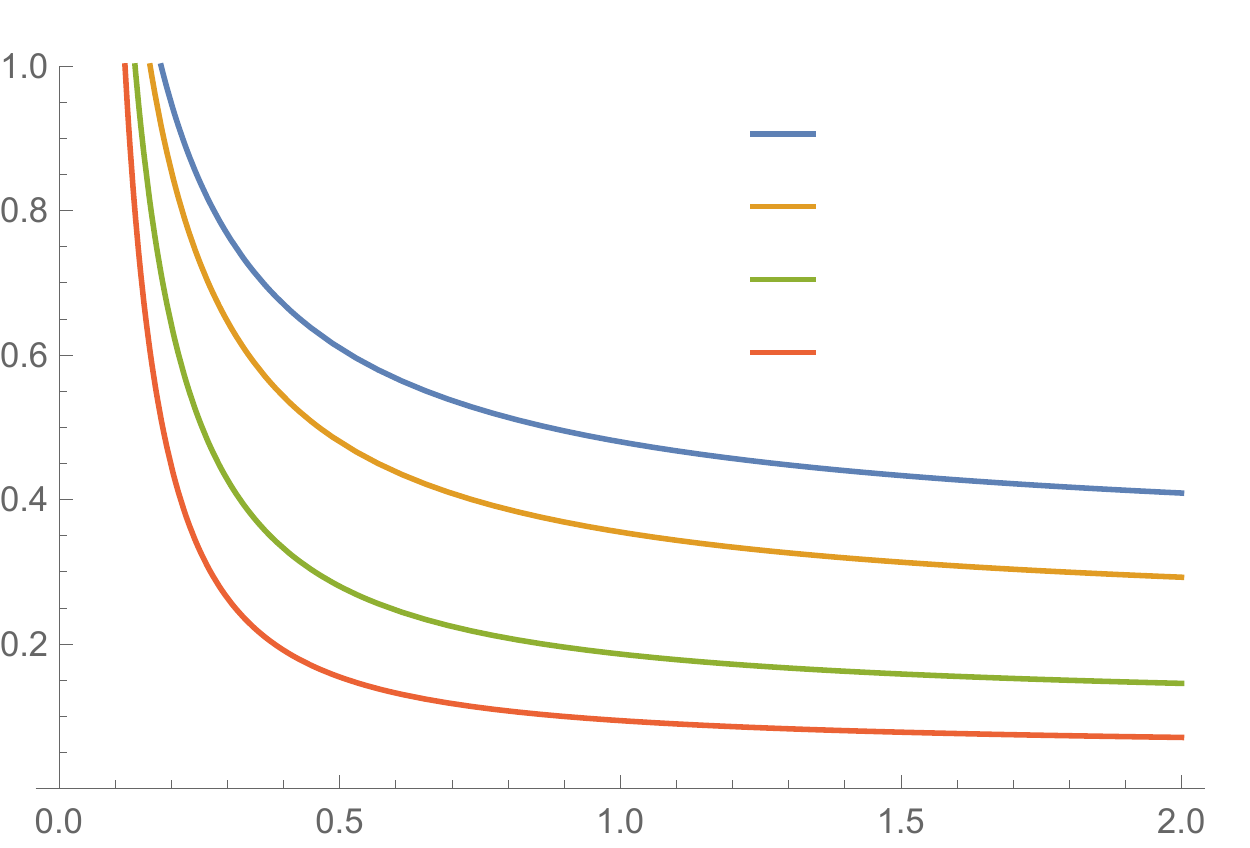}}
		\put(0.1,3.5){\includegraphics[width=0.46\textwidth]{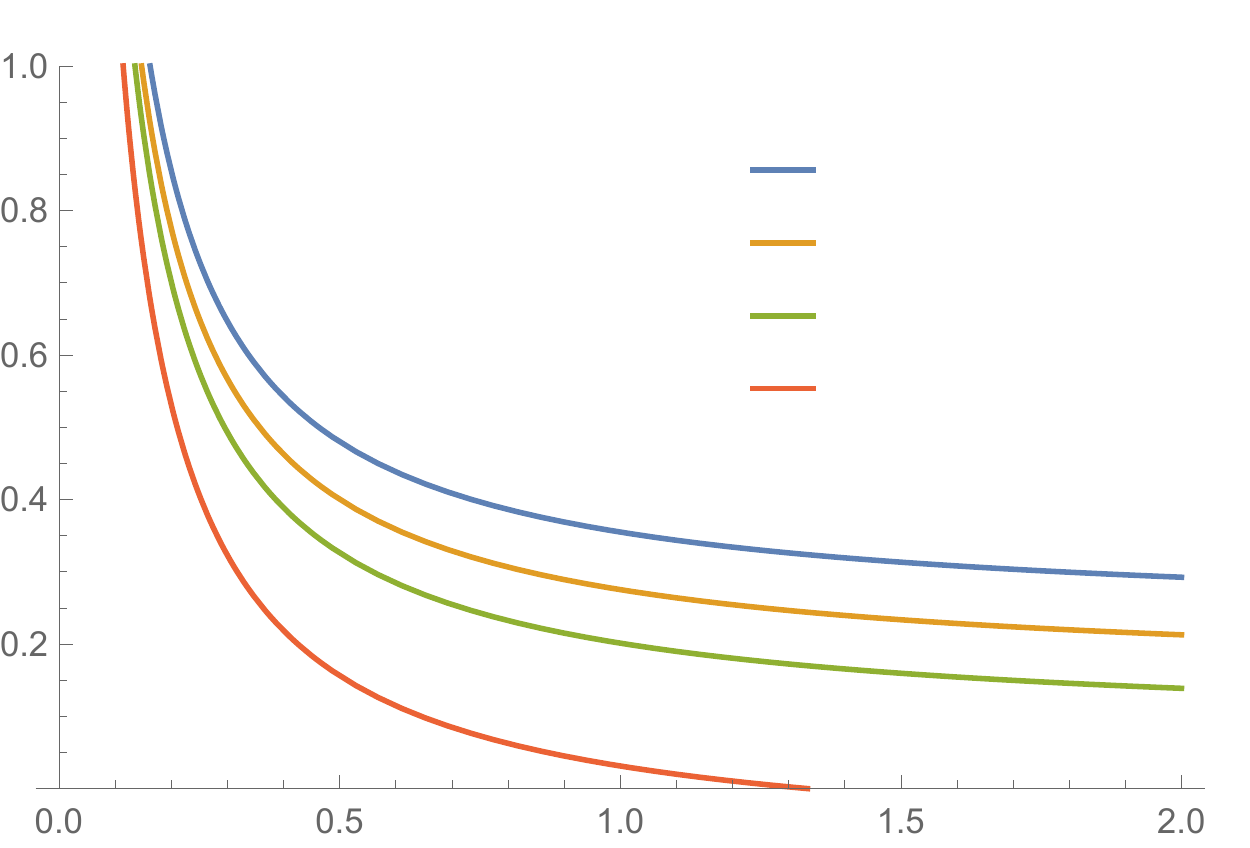}}
		\put(5.2,0){\includegraphics[width=0.46\textwidth]{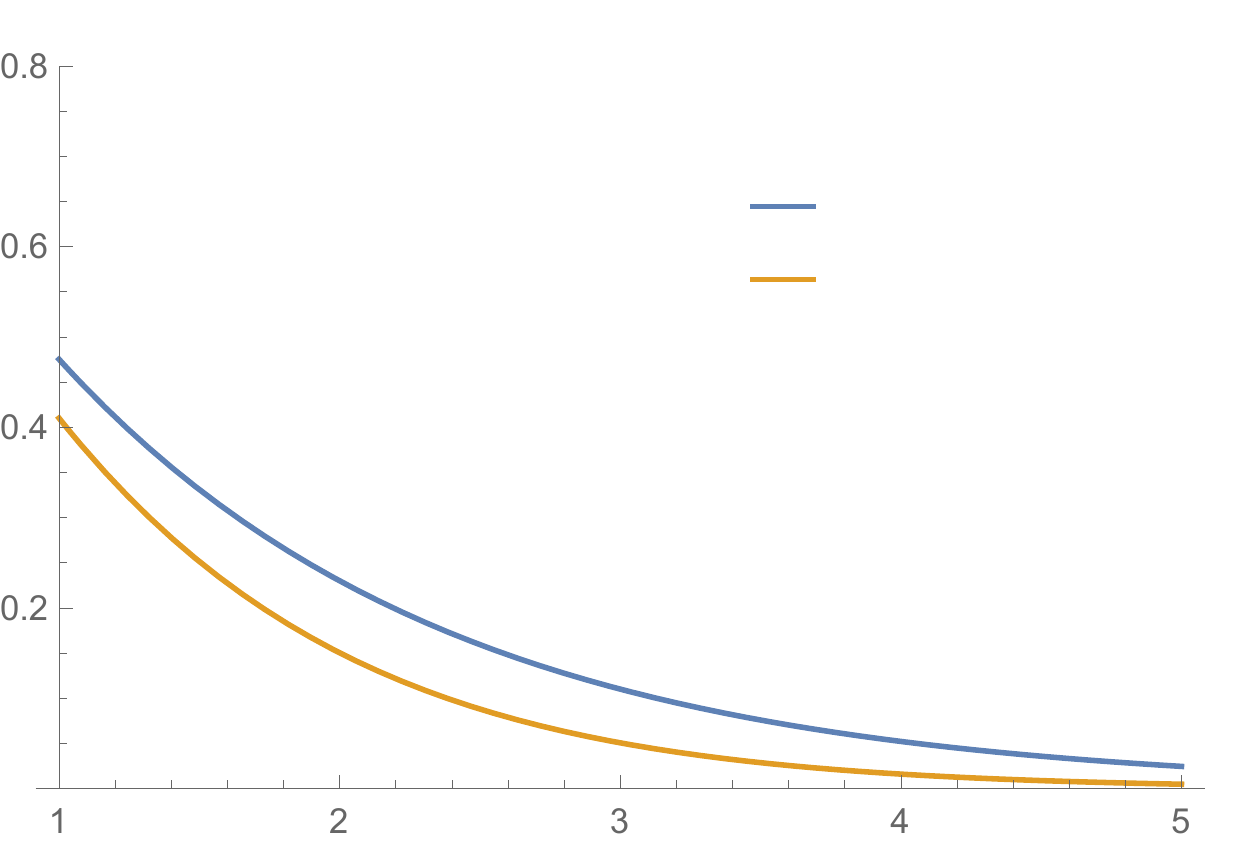}}
		\put(5.2,3.5){\includegraphics[width=0.46\textwidth]{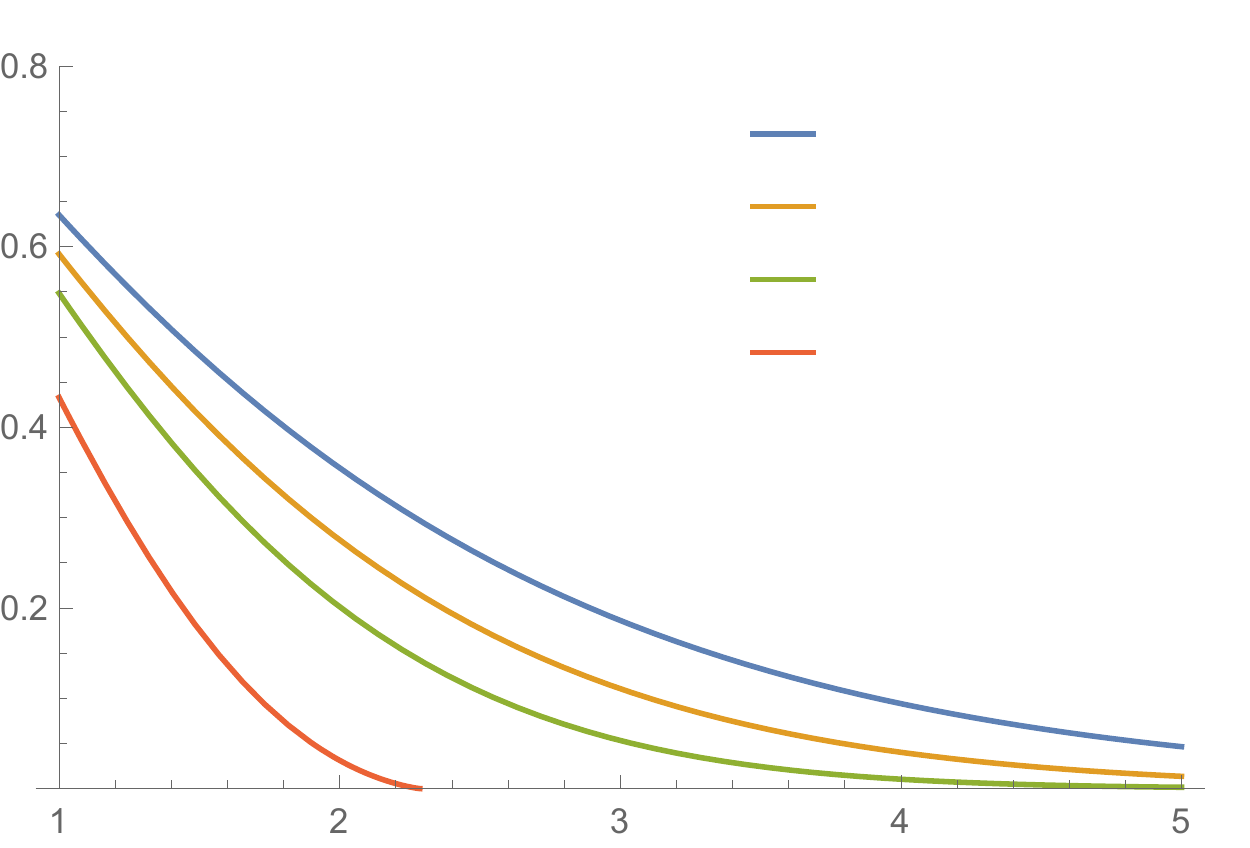}}
     \put(0.7,6.9){\underline{$z=2$}}
		\put(0.25,6.5){${\scriptstyle S_n}$}
		\put(4.63,3.68){${\scriptstyle n}$}
		\put(3.22,5.94){${\scriptstyle d_\theta=2, \, \Phi=1.1}$}
		\put(3.22,5.67){${\scriptstyle d_\theta=2, \, \Phi=1.025}$}
		\put(3.22,5.395){${\scriptstyle d_\theta=2, \, \Phi=0.95}$}
		\put(3.22,5.13){${\scriptstyle d_\theta=2, \, \Phi=0.75}$}
		\put(0.25,3.01){${\scriptstyle S_n}$}
		\put(4.63,0.18){${\scriptstyle n}$}
		\put(3.22,2.572){${\scriptstyle d_\theta=1.5, \, \Phi=1.1}$}
		\put(3.22,2.298){${\scriptstyle d_\theta=2, \, \Phi=1.1}$}
		\put(3.22,2.032){${\scriptstyle d_\theta=3, \, \Phi=1.1}$}
		\put(3.22,1.762){${\scriptstyle d_\theta=4, \, \Phi=1.1}$}
		\put(5.35,6.5){${\scriptstyle S_1}$}
		\put(9.73,3.68){${\scriptstyle d_\theta}$}
		\put(8.31,6.072){${\scriptstyle n=1, \, \Phi=1.1}$}
		\put(8.31,5.802){${\scriptstyle n=1, \, \Phi=1.025}$}
		\put(8.31,5.540){${\scriptstyle n=1, \, \Phi=0.95}$}
		\put(8.31,5.263){${\scriptstyle n=1, \, \Phi=0.75}$}
		\put(5.31,3.01){${\scriptstyle S_\infty}$}
		\put(9.73,0.18){${\scriptstyle d_\theta}$}
		\put(8.31,2.307){${\scriptstyle n=\infty, \, \Phi=1.1}$}
		\put(8.31,2.033){${\scriptstyle n=\infty, \, \Phi=1.025}$}
   \end{picture}
   \caption{Plot of R\'enyi entropies with Lifshitz parameter $z=2$. The normalization of the parameters are the same as the previous plots have. The upper left panel shows that R\'enyi entropy can be negative if $\Phi^2<\Phi^2_c$, but this is already excluded if one takes into account the constraint $S_{\rm{BH}}(T/n)\geq 0$. On the upper right panel a minimal value of $S_1$ appears for the plot with potential $\Phi=0.75$. The reason behind this is that there is no positive temperature corresponding to the parameter choice. The lower right panel again indicates that the largest eigenvalue $\lambda_1$ goes to one.}
   \label{fig:Plotz2} 
\end{figure}
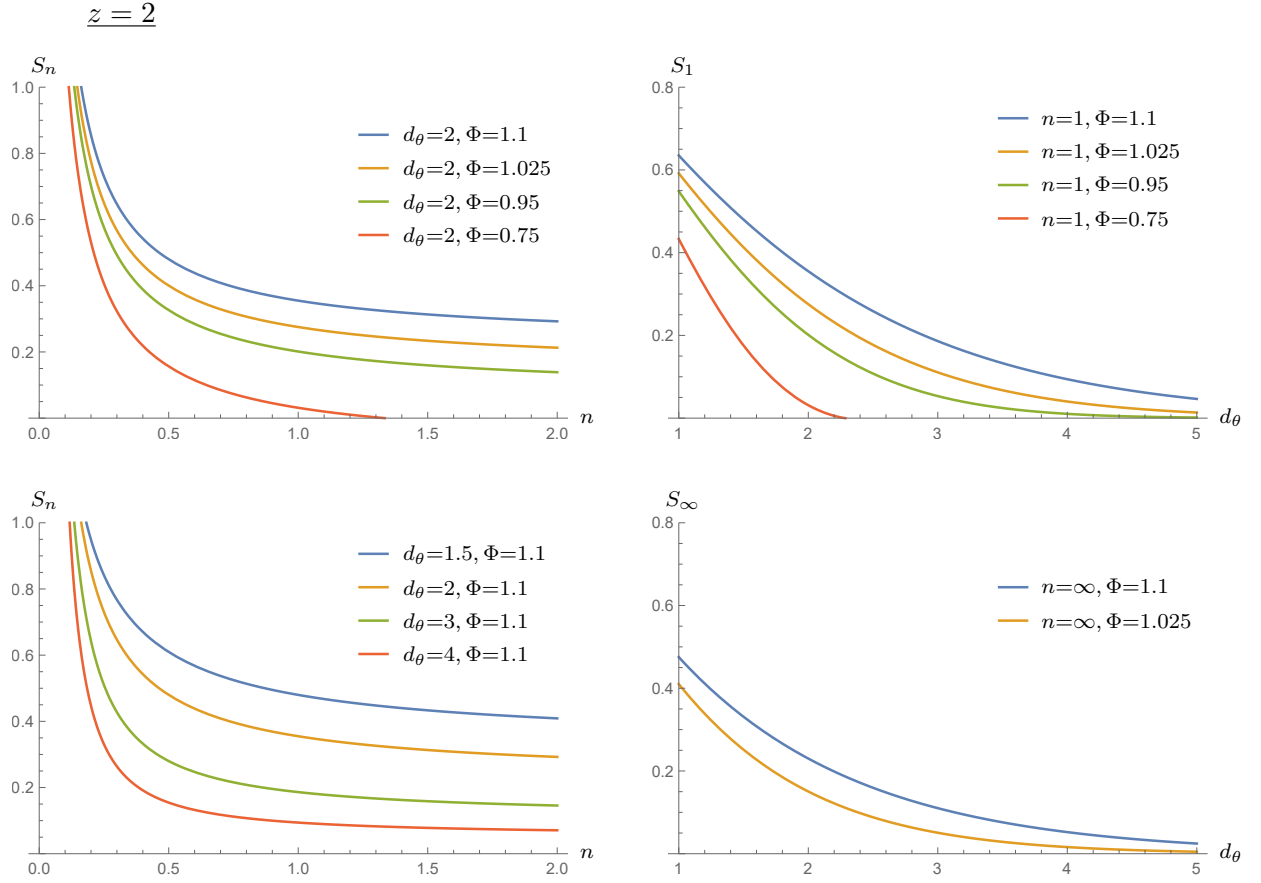

{\underline{\sl The case $z=2$.} \ }
Lifshitz scaling with $z=2$ appears in specific condensed matter structures such that magnetic materials and liquid crystals \cite{Kachru2008}, and it was used to study superconducting fluctuations in various context (see e.g. \cite{Chemissany2012,Bu2012,Fan2013,Lu2013,Gursel2019}).

The algebraic equation \eqref{eq:TperneqTxn} is second order and the $n$-sheeted horizon $x_n$ has the solution
\be
x_n \, = \, \sqrt{\frac{\frac{4\pi \ell T}{n}-(d-\theta)c^2\ell^{-2}(\Phi_c^2-\Phi^2)}{d-\theta+2}} \ .
\ee
It exists for all $n$ if $k=0,-1$ or $k=1$ with $\Phi^2>\Phi_c^2$, but yields the maximal value 
\be
n_{\text{max}} \, = \,\frac{4\pi \ell^3 T}{(d-\theta)c^2(\Phi_c^2-\Phi^2)} 
\ee 
for spherical topology if $\Phi_c^2>\Phi^2$. Then the qualitative behavior of R\'enyi entropy with this solution agrees with the first and third parameter cases discussed in the previous subsection depending on the value of the electric potential. The maximal R\'enyi parameter we calculated can be also derived from \eqref{eq:nmaxBH} for $z=2$. We illustrate the calculation with plots in Figure~\ref{fig:Plotz2}. It shows in some explicit cases how the R\'enyi entropy goes to a finite value for large $n$ or fails to be positive below the former upper bound $n_{\text{max}}$. The latter case is an interesting feature, which we only found for $z=2$, because the positivity of R\'enyi entropy is a more fundamental constraint than the one related to the thermodynamic stability. It originates from the second R\'enyi inequality \eqref{eq:RenyiIneq}, which ensures the positivity of $S(T/n)$ (see \eqref{eq:condEntrSpecHeatPos}) and thus the positivity of $S_n$. In \cite{Dong2016} they showed that the quantity $n^2\partial_n (\frac{n-1}{n} S_n)$ computes the area of a dual cosmic brane with a conical deficit angle $2\pi\frac{n-1}{n}$, so the second inequality expresses that a dual bulk cosmic brane geometry has a positive area.  Hence our computation of R\'enyi entropy shows that there are cases with $z=2$ and spherical topologies when the dual cosmic brane picture can be problematic.

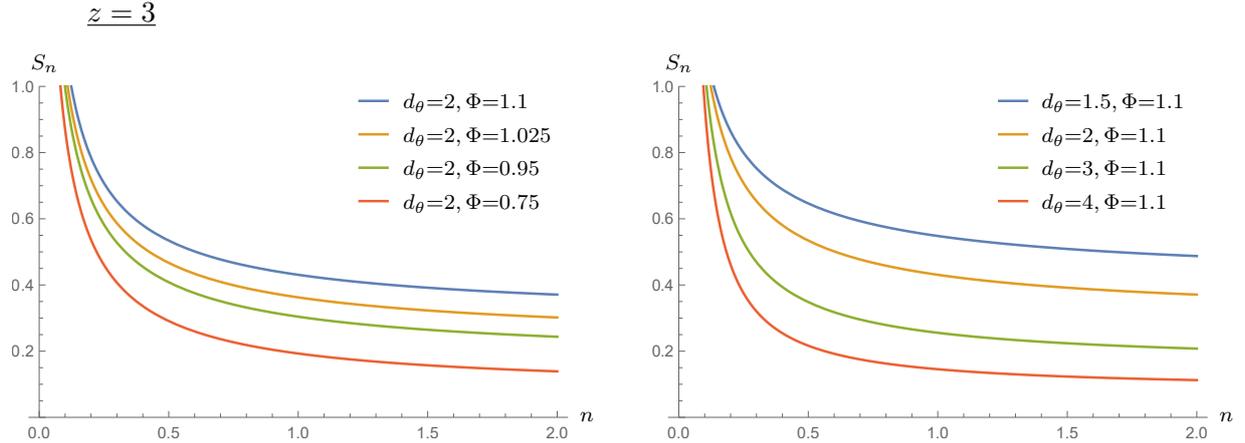
\begin{figure}[htb]
   \centering
  \setlength{\unitlength}{0.1\textwidth}
   \begin{picture}(10,3.75)
     \put(0.1,0){\includegraphics[width=0.46\textwidth]{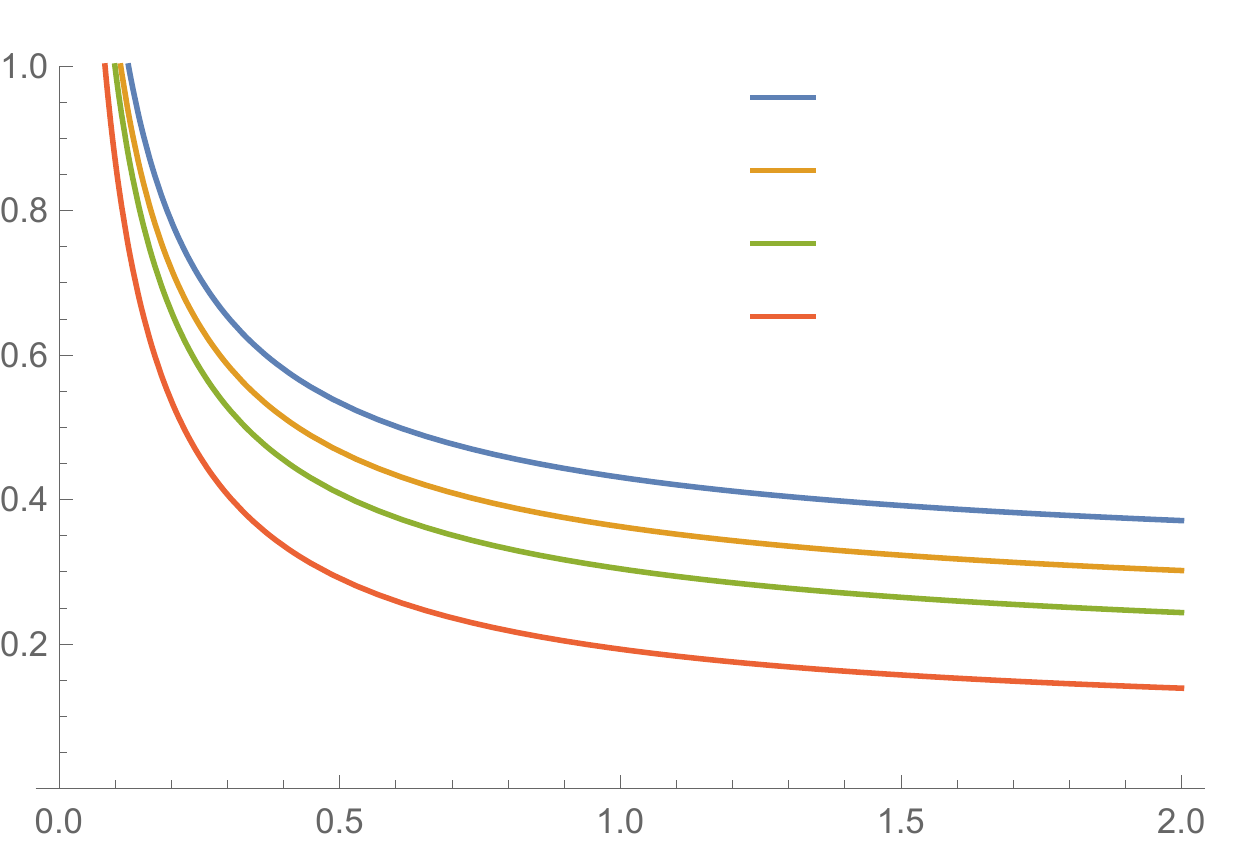}}
		\put(5.2,0){\includegraphics[width=0.46\textwidth]{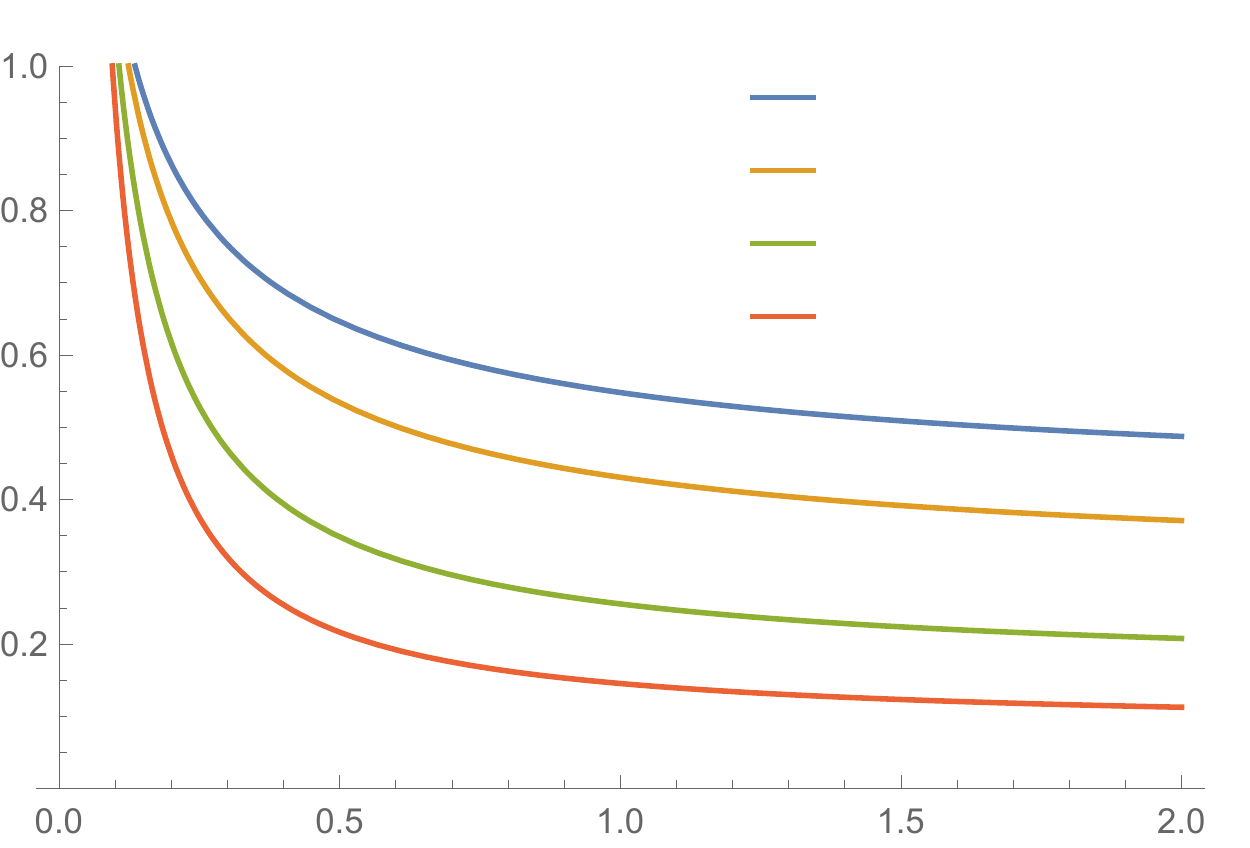}}
     \put(0.7,3.4){\underline{$z=3$}}
			\put(0.25,3.01){${\scriptstyle S_n}$}
		\put(4.63,0.18){${\scriptstyle n}$}
		\put(3.22,2.702){${\scriptstyle d_\theta=2, \, \Phi=1.1}$}
		\put(3.22,2.428){${\scriptstyle d_\theta=2, \, \Phi=1.025}$}
		\put(3.22,2.162){${\scriptstyle d_\theta=2, \, \Phi=0.95}$}
		\put(3.22,1.892){${\scriptstyle d_\theta=2, \, \Phi=0.75}$}
			\put(5.31,3.01){${\scriptstyle S_n}$}
		\put(9.73,0.18){${\scriptstyle n}$}
		\put(8.31,2.702){${\scriptstyle d_\theta=1.5, \, \Phi=1.1 }$}
		\put(8.31,2.428){${\scriptstyle d_\theta=2, \, \Phi=1.1}$}
		\put(8.31,2.162){${\scriptstyle d_\theta=3, \, \Phi=1.1}$}
		\put(8.31,1.892){${\scriptstyle d_\theta=4, \, \Phi=1.1}$}
   \end{picture}
   \caption{Plots of R\'enyi entropies with Lifshitz parameter $z=3$. Different $\Phi$ and $d_\theta$ parameters were used to illustrate its behavior qualitatively. We used the same normalization of other parameters as before. One can see that R\'enyi entropy exists for all R\'enyi parameter.}
   \label{fig:Plotz3} 
\end{figure}

{\underline{\sl The case $z=3$.} \ }
Lifshitz scaling with $z=3$ was studied as geometric background of holographic superconductors \cite{Lu2013} and also appeared in the context of null normalizable deformation of AdS${}_5$ \cite{Narayan2012}. 
The equation \eqref{eq:TperneqTxn} is a third order polynomial equation, and there is only one real and positive solution $x_n$ for every $n$ regardless of the electric potential. Thus the possible values of the R\'enyi parameter ranges from zero to infinity, which coincides with that the dual cosmic brane picture is expected to be valid for $z>2$. Instead of giving an expression for the root $x_n$, we illustrate the R\'enyi entropy as a function of the parameters with some examples in Figure~\ref{fig:Plotz3}.  Since $z=3$ belongs to the second and fourth cases in the previous subsection, we can conclude from the discussion there that $x_\infty$ goes to zero value for spherical topologies with $\Phi<\Phi_c$ and a finite value otherwise. Hence the R\'enyi entropy goes to a finite value if $n$ approaches infinity. The plots also illustrate how the R\'enyi entropies decreases with the electric potential or the effective dimension $d_\theta$ as well as for other values of $z$.

\begin{figure}[htb]
   \centering
  \setlength{\unitlength}{0.1\textwidth}
   \begin{picture}(10,7.2)
     \put(0.1,0){\includegraphics[width=0.46\textwidth]{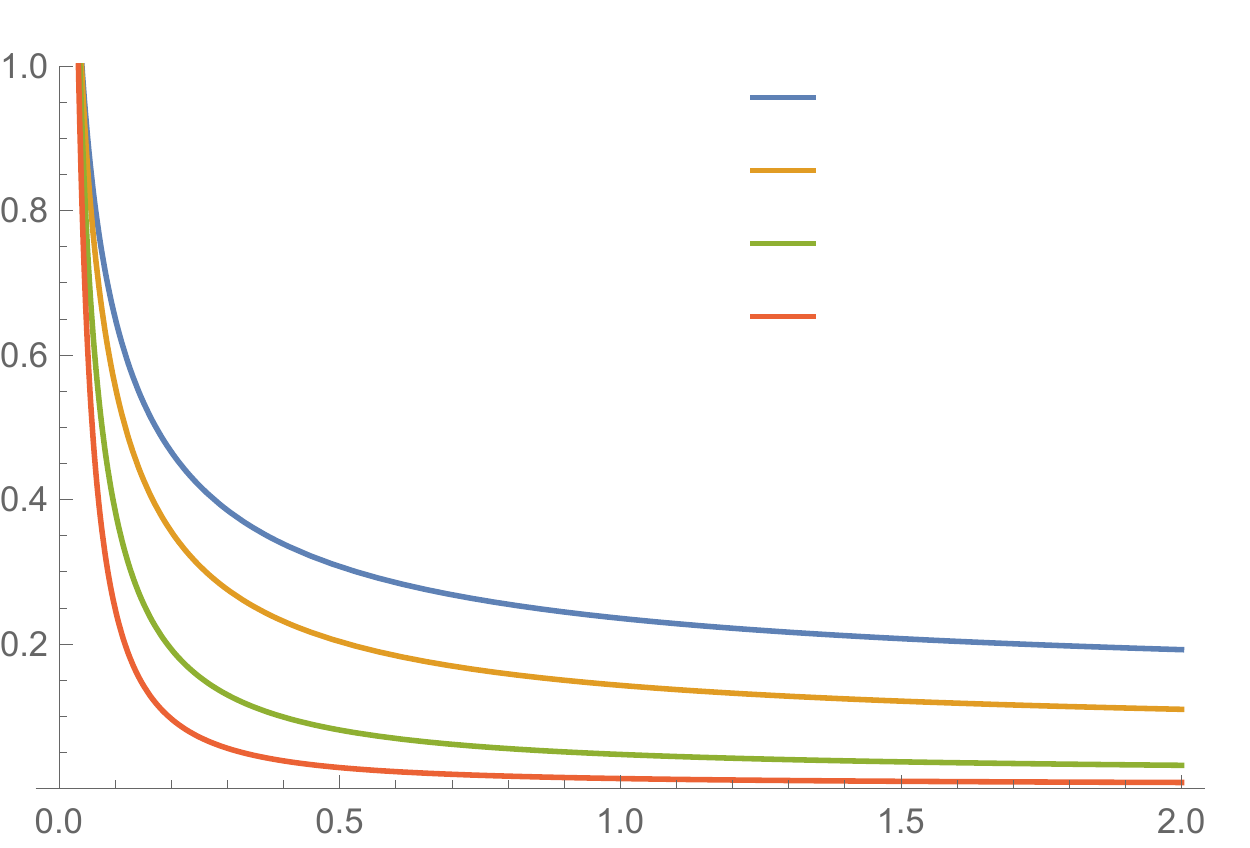}}
		\put(0.1,3.5){\includegraphics[width=0.46\textwidth]{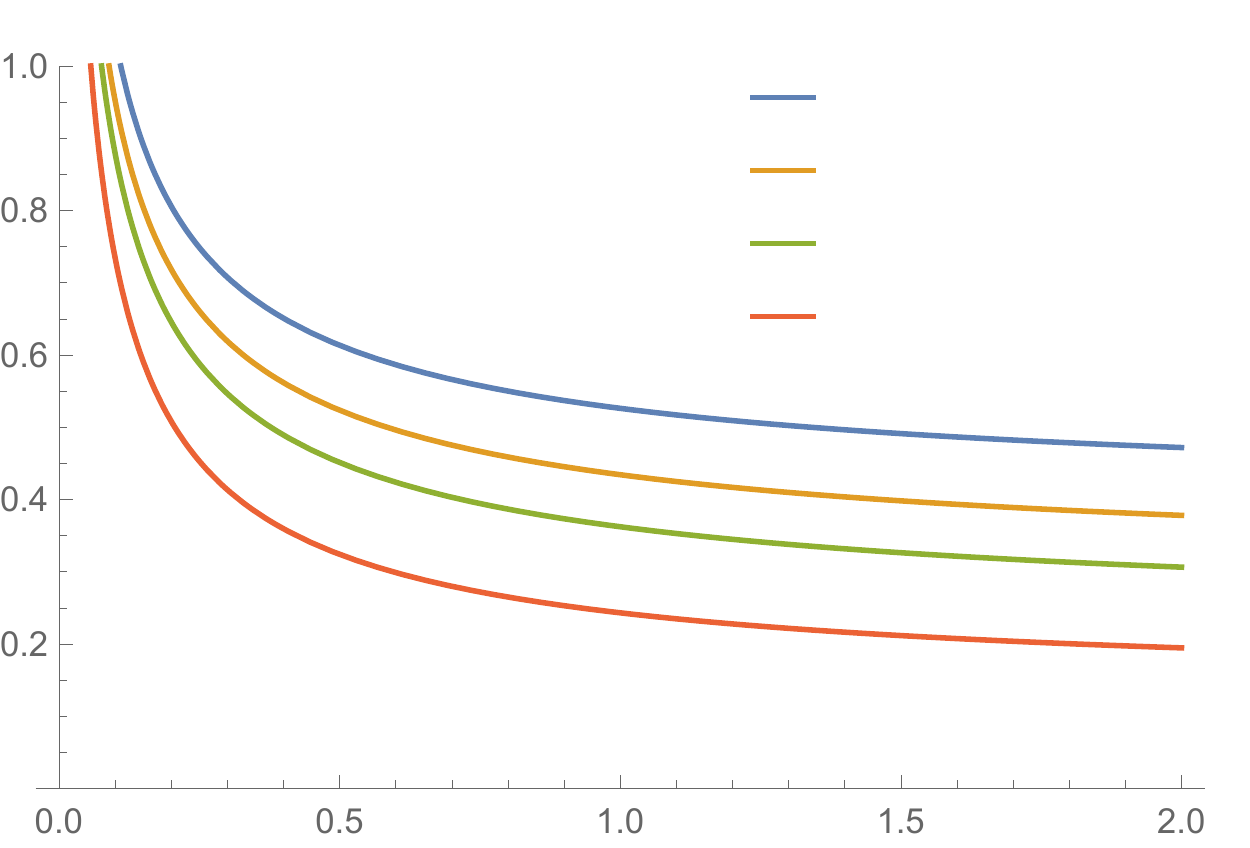}}
		\put(5.2,0){\includegraphics[width=0.46\textwidth]{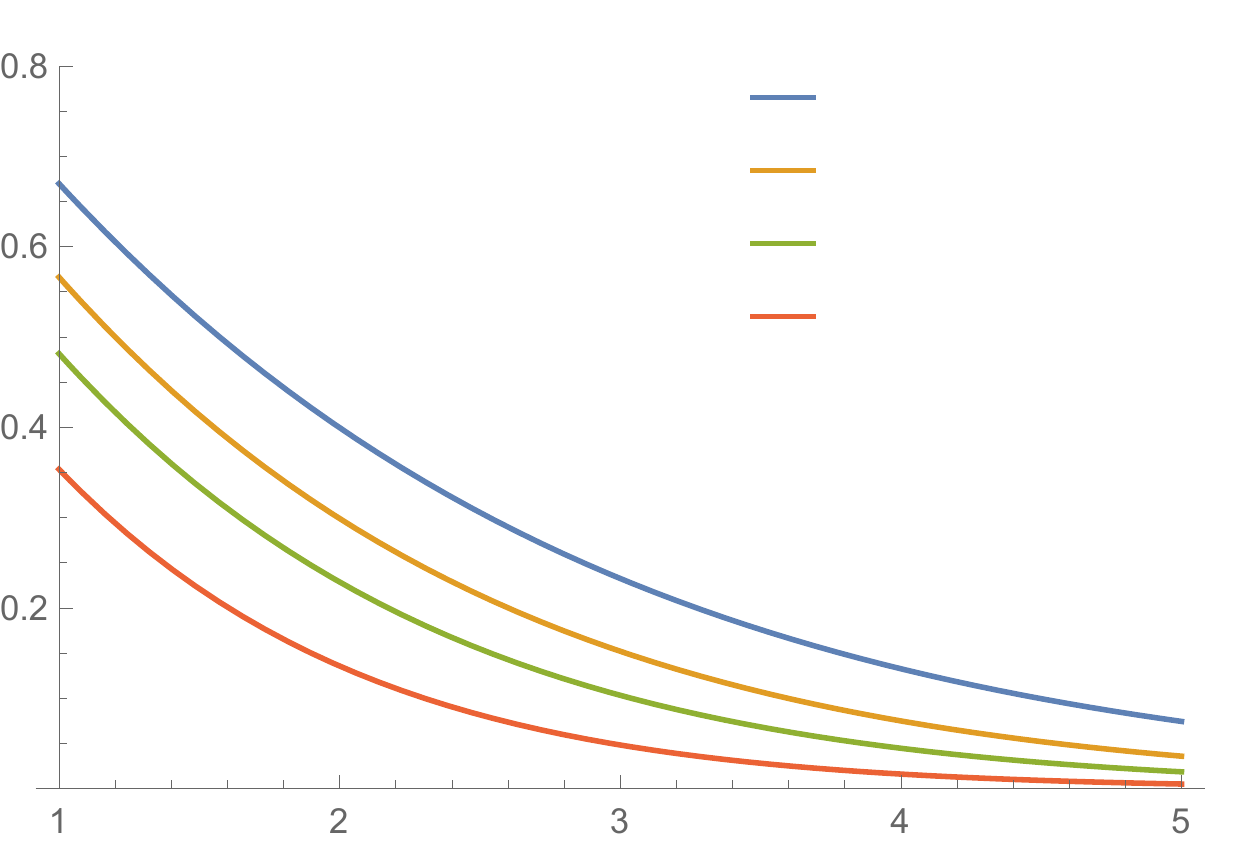}}
		\put(5.2,3.5){\includegraphics[width=0.46\textwidth]{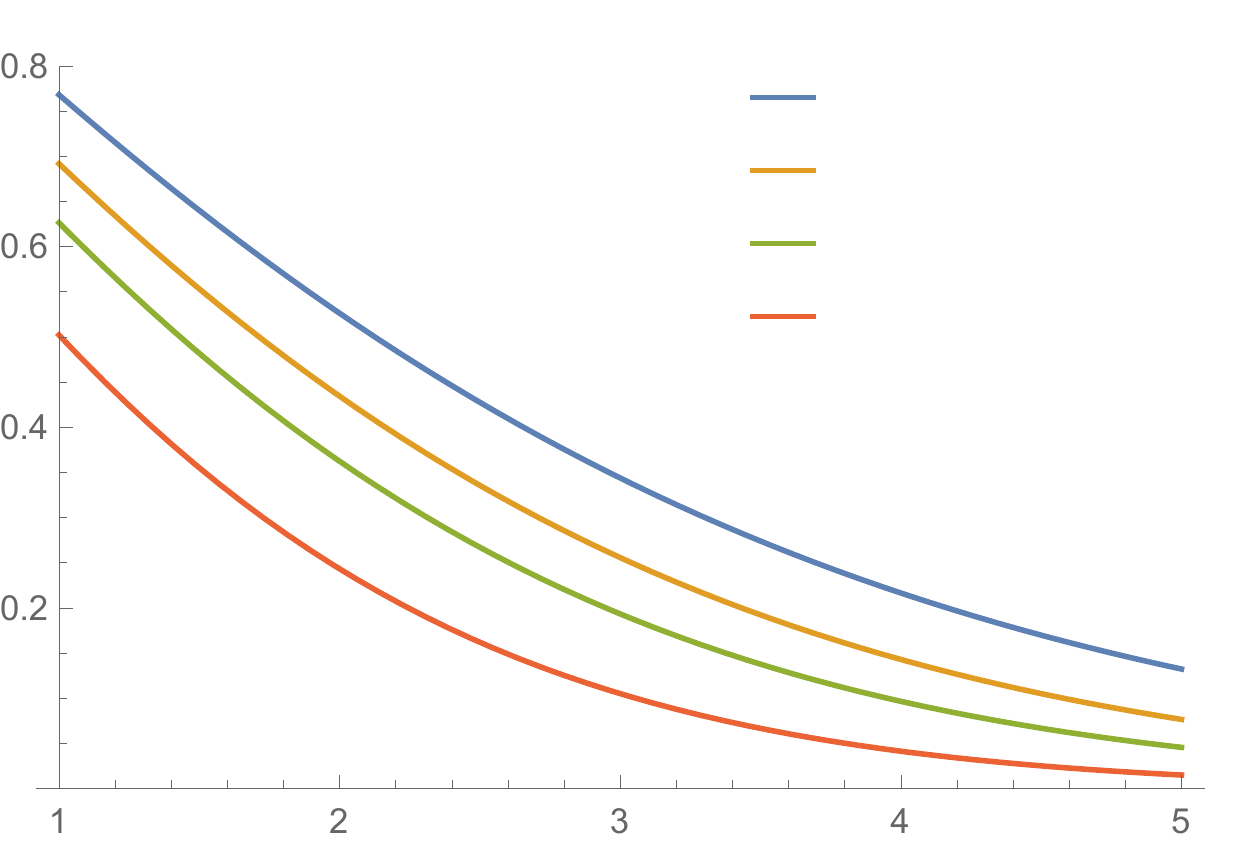}}
     \put(0.7,6.9){\underline{$z=4$}}
		\put(0.25,6.5){${\scriptstyle S_n}$}
		\put(4.63,3.68){${\scriptstyle n}$}
		\put(3.22,6.205){${\scriptstyle d_\theta=2, \, \Phi=1.1}$}
		\put(3.22,5.935){${\scriptstyle d_\theta=2, \, \Phi=1.025}$}
		\put(3.22,5.66){${\scriptstyle d_\theta=2, \, \Phi=0.95}$}
		\put(3.22,5.395){${\scriptstyle d_\theta=2, \, \Phi=0.75}$}
		\put(0.25,3.01){${\scriptstyle S_n}$}
		\put(4.63,0.18){${\scriptstyle n}$}
		\put(3.22,2.704){${\scriptstyle d_\theta=1.5, \, \Phi=1.1}$}
		\put(3.22,2.433){${\scriptstyle d_\theta=2, \, \Phi=1.1}$}
		\put(3.22,2.164){${\scriptstyle d_\theta=3, \, \Phi=1.1}$}
		\put(3.22,1.894){${\scriptstyle d_\theta=4, \, \Phi=1.1}$}
		\put(5.35,6.5){${\scriptstyle S_1}$}
		\put(9.73,3.547){${\scriptstyle d_\theta}$}
		\put(8.31,6.204){${\scriptstyle n=1, \, \Phi=1.1}$}
		\put(8.31,5.934){${\scriptstyle n=1, \, \Phi=1.025}$}
		\put(8.31,5.672){${\scriptstyle n=1, \, \Phi=0.95}$}
		\put(8.31,5.395){${\scriptstyle n=1, \, \Phi=0.75}$}
		\put(5.31,3.01){${\scriptstyle S_\infty}$}
		\put(9.73,0.18){${\scriptstyle d_\theta}$}
		\put(8.31,2.704){${\scriptstyle n=\infty, \, \Phi=1.1}$}
		\put(8.31,2.439){${\scriptstyle n=\infty, \, \Phi=1.025}$}
		\put(8.31,2.165){${\scriptstyle n=\infty, \, \Phi=0.95}$}
		\put(8.31,1.896){${\scriptstyle n=\infty, \, \Phi=0.75}$}
   \end{picture}
   \caption{R\'enyi entropies with Lifshitz parameter $z=4$ and different fixed values of $\Phi$ and $d_\theta$. The normalization of other parameters are the same as before. The plots shows that all R\'enyi entropy exists for parameters zero to infinity, hence all limits are meaningful, and suggest that the largest eigenvalue is again goes to one.}
   \label{fig:Plotz4} 
\end{figure}
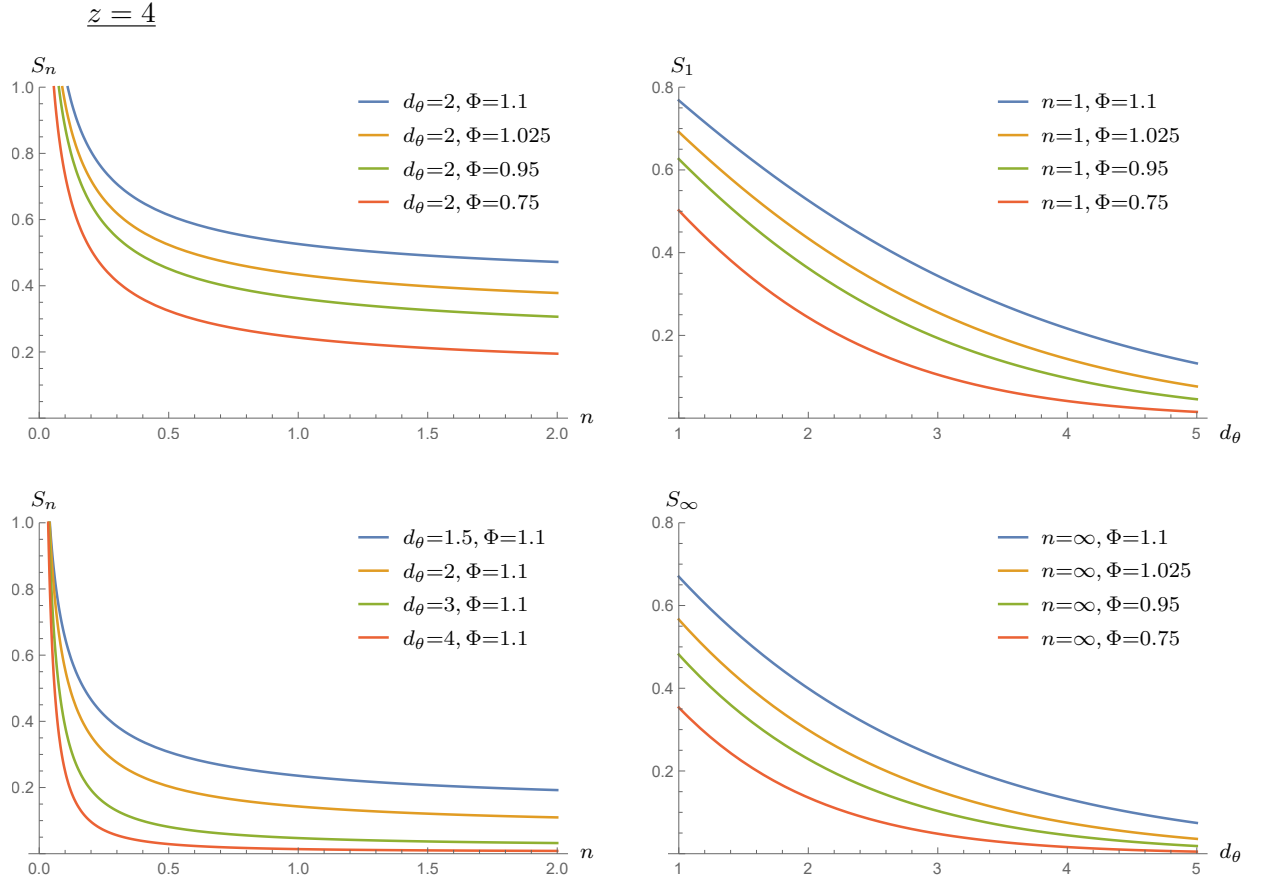

{\underline{\sl The case $z=4$.} \ }
One relevant root exists for all values of $n$, which is
\be
x_n \, = \, \sqrt{\frac{\sqrt{(d-\theta+3)^2 c^4 \ell^{-4}(\Phi_c^2-\Phi^2)^2+\frac{16\pi\ell T}{n}(d-\theta+4)}-(d-\theta+3)c^2 \ell^{-2}(\Phi_c^2-\Phi^2)}{2(d-\theta+4)}} \ .
\ee
Similarly to the $z=3$ case $x_\infty$ goes to zero value for spherical topologies and $\Phi < \Phi_c$, and a finite value otherwise. Thus R\'enyi entropy exists for all values of the R\'enyi parameter $n$, and there are no restrictions by the inequalities. We show some illustrative plots in Figure~\ref{fig:Plotz4}. One can see that similarly to lower value of $z$-s, the R\'enyi entropy decreases with the electric potential and effective dimension.  

\subsection{Large $d_\theta$ limit}

We have seen that the spectrum of the cases studied above have an interesting feature in large $d_\theta=d-\theta$ limit. If the R\'enyi entropy exists, it seems to approach to zero, which we study here in more detail. We first choose the Newton constant such that $\frac{GT}{\ell^{d_\theta} \omega_{k,d}} = \mathcal{O}(1)$. There are two possibilities for temperature: if we choose it to be finite with respect to $d_\theta$, then the solution of the horizon radius for all R\'enyi parameter is the same in large $d_\theta$, which is\footnote{The case, when the R\'enyi parameter is such small, that it is in the order of $1/d_\theta$, is excluded in the present discussion.}
\be \label{eq:BounddthetaInf}
\bar{x}\, = \, \sqrt{\frac{c^2(\Phi^2-\Phi^2_c)}{\ell^2}}
\ee
for $\Phi_c^2\leq\Phi^2$, and zero if $\Phi^2 <\Phi_c^2$ ($k=1$). Then the R\'enyi entropy is zero, since the horizon $x_n$ is independent of $n$. Also $S_\infty$ goes to zero, which is equals to $(E_1-G)/T$, where $E_1$ is the ground state energy of the modular Hamiltonian $H-\Phi Q$. It follows that $E_1=G$ in this limit.

We can also choose $T$ to run with $d_\theta$ linearly. In this case the R\'enyi entropy is not zero, but goes to an $n$-dependent value. If $n<1$, then $G$ approaches $-\frac{n}{n-1}\frac{G_n}{T}$ at large $d_\theta$, while with parameter $n>1$, it goes to $-\frac{n}{n-1}\frac{G}{T}$. Hence $S_\infty$ has the finite value $-G/T$, and it yields zero ground state energy $E_1=0$ for the modular Hamiltonian.

\begin{figure}[hbt]
   \centering
  \setlength{\unitlength}{0.1\textwidth}
   \begin{picture}(10,3.85)
     \put(0.1,0){\includegraphics[width=0.46\textwidth]{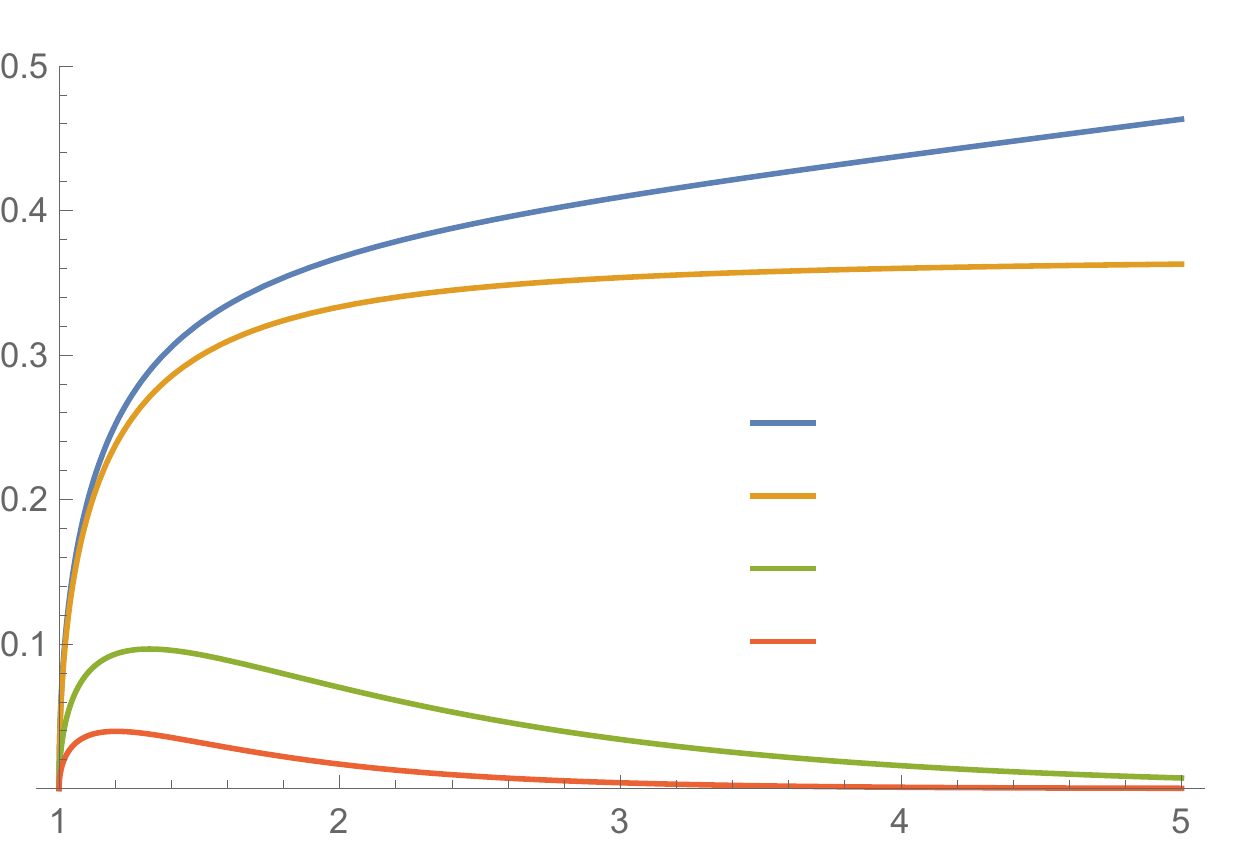}}
		\put(5.2,0){\includegraphics[width=0.46\textwidth]{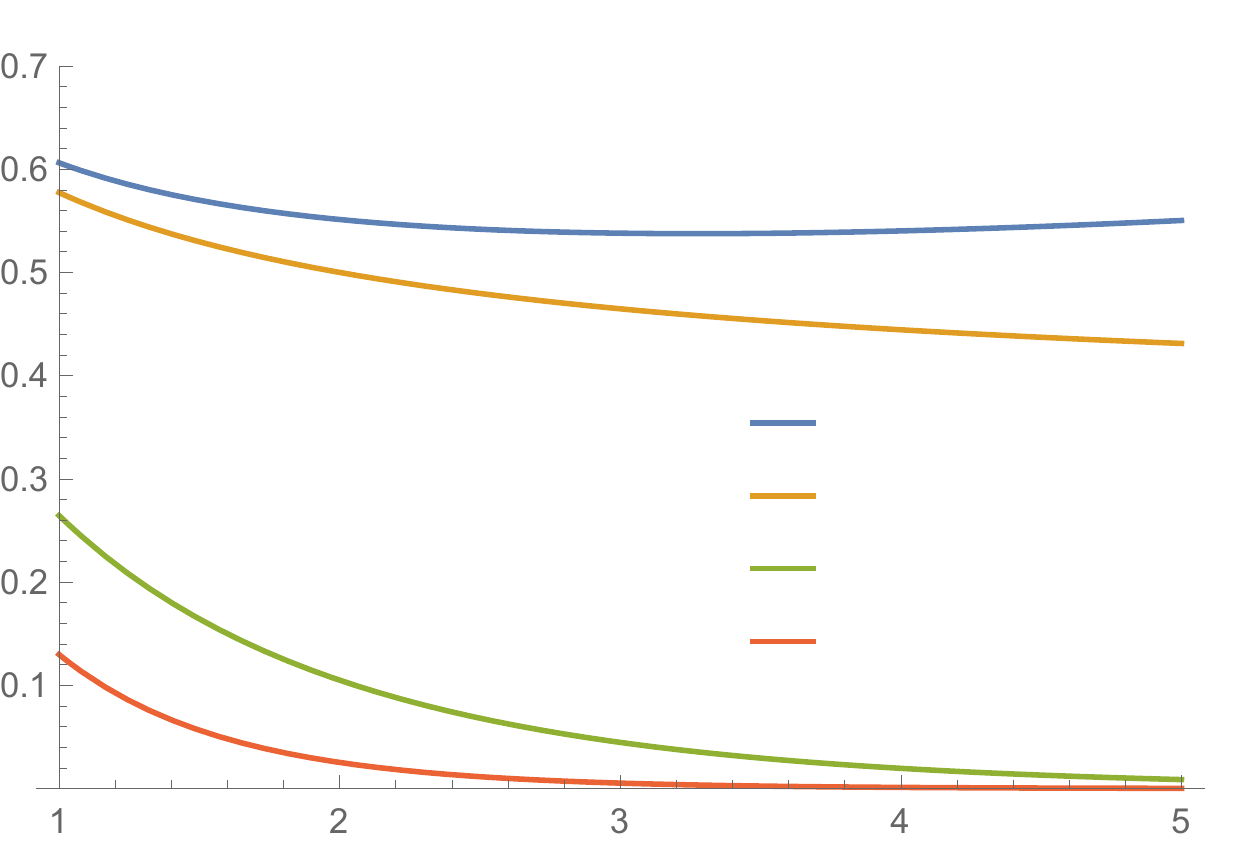}}
     \put(0.7,3.4){\underline{$z=1$}}
		  \put(5.7,3.4){\underline{$z=2$}}
				\put(0.25,3.01){${\scriptstyle S(0)}$}
		\put(4.63,0.18){${\scriptstyle d_\theta}$}
		\put(3.22,1.506){${\scriptstyle \Phi=1.45}$}
		\put(3.22,1.235){${\scriptstyle \Phi=\sqrt{2}}$}
		\put(3.22,0.967){${\scriptstyle \Phi=1.1}$}
		\put(3.22,0.697){${\scriptstyle \Phi=1.025}$}
			\put(5.31,3.01){${\scriptstyle S(0)}$}
		\put(9.73,0.18){${\scriptstyle d_\theta}$}
		\put(8.31,1.506){${\scriptstyle \Phi=1.45 }$}
		\put(8.31,1.235){${\scriptstyle \Phi=\sqrt{2}}$}
		\put(8.31,0.967){${\scriptstyle \Phi=1.1}$}
		\put(8.31,0.697){${\scriptstyle \Phi=1.025}$}
   \end{picture}
   \caption{Plot of Bekenstein-Hawking entropies at zero temerature, with $z=1$ on the left and $z=2$ on the right. They go towards zero if $\Phi < \sqrt{2}$ with our parameter choice, and blow up if $\sqrt{2} < \Phi$. This suggests that the ground state degeneracy of the modular Hamiltonian approaches one below the bound \eqref{eq:BounddthetaInf}, while blows up above it.}
   \label{fig:SBH0z12} 
\end{figure}
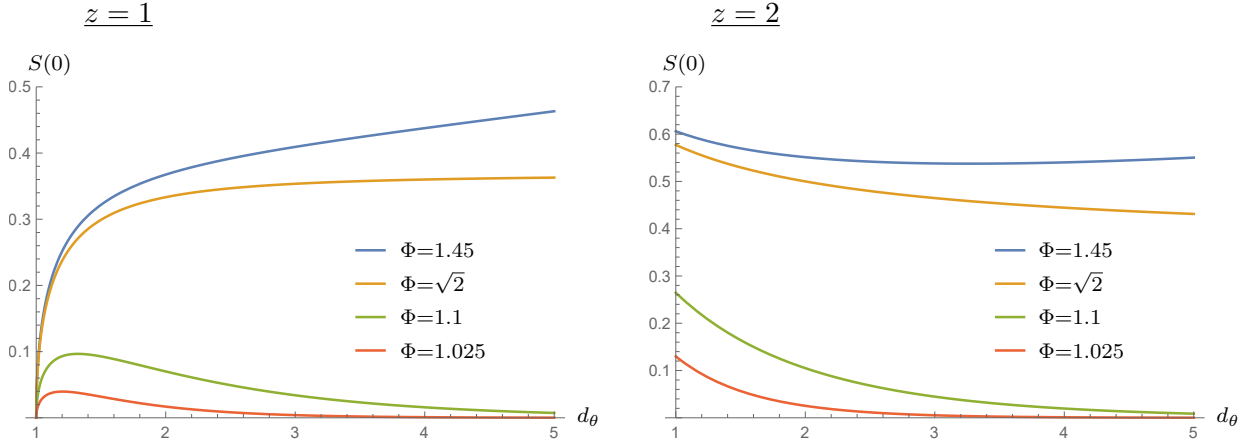

Another interesting quantity, which carries information about the spectrum is the Bekenstein-Hawking entropy at zero temperature (see e.g. \cite{Myers2011}). It gives the degeneracy of the largest eigenvalue $\lambda_1$ of the density matrix $\rho$ by
\be
\lim_{n\rightarrow \infty} S(T/n) = \log d(\lambda_1) \ .
\ee
We show some examples for $z=1,2$ in Figure~\ref{fig:SBH0z12} to illustrate this quantity. The entropy $S(0)$ at large $d_\theta$ and potential $\Phi_c^2\leq\Phi^2$ goes to zero, if $\bar{x}$ given in \eqref{eq:BounddthetaInf} is below one, and blows up if it is above. For potential $\Phi^2<\Phi_c^2$ the ground state entropy $S(0)$ always goes to zero, because the horizon radius corresponding to $T=0$ is zero. To summarize the two cases, the R\'enyi entropy is zero, if $\Phi^2\leq \Phi_c^2 +\ell^2/c^2$. This argument suggests that the degeneracy of ground state of the dual modular Hamiltonian goes to one or blows up depending on the electric potential. If it is smaller than $\Phi_c^2 +\ell^2/c^2$, the spectrum of the dual theory seems to be simplifying in the large $d_\theta$ limit, at least it suggests that the ground state is unique in the classical limit we studied.

The aspects of $d_\theta$ limit we studied above can be interpreted by what happens to the background geometry. The gravitational potential terms in the blackening factor \eqref{eq:BlackeningFactorGenDimZTh} smear out at finite distance from the horizon, only the term corresponding to the non-trivial topologies remains, that is
\be
f(r) \ \longrightarrow \ 1 \, + \, \frac{c^2\Phi^2_c}{r^2} \, + \, \mathcal{O}\big(\left(r_h/r\right)^{d_\theta}\big) \ ,
\ee 
where we kept $\Phi_c$ as well as $\Phi$ at order one.

Now we separate our discussion between $\theta \rightarrow -\infty$ and $d \rightarrow \infty$, which are both realization of large $d_\theta$. When $\theta$ approaches negative infinity, we use the substitution of coordinates $\rho=(r/r_F)^{-\theta/d}$. Since $r_F$ has the role of an upper cutoff of the theory, we assume $r<r_F$. Then the metric \eqref{eq:HypeLifmetricGenDim} at finite distance from the horizon takes the form
\be
ds^2 \, = \, \rho^2 \left(- \left(\frac{r_F}{\ell}\right)^{2z} K \, dt^2 \, + \, \frac{d^2 \ell^2}{\theta^2 K} \, \frac{d \rho^2}{\rho^2} \, + \, r_F^2 \, d \Omega^2_{k,d} \right) \ , \qquad \quad K=1+\frac{c^2 \Phi_c^2}{r_F^2} \ ,
\ee 
which shows that the singularity is smeared out. In order to stay in the holographic regime we need to keep the Newton constant $G\propto\ell^{d_\theta}\omega_{k,d}/T$ small.  Since we took $c^2\Phi_c^2$ to be finite, then $\ell=\mathcal{O}(\theta)$, which yields that we need to choose $\ell < 1$, if $T=\mathcal{O}(1)$, and $\ell \leq 1$, if $T=\mathcal{O}(d_\theta)$. Then $G$ goes to zero in the $\theta\rightarrow -\infty$ limit.

If we realize large $d_\theta$ with large dimension $d$, we encounter a similar problem to that studied in~\cite{Emparan2013,Emparan2015,Bhattacharyya2015,Emparan2015a,Bhattacharyya2015a} for mostly general relativity and AdS spacetime. The metric becomes asymptotically Lifshitz, that is
\be
ds^2 \, = \, \left(\frac{r}{\ell}\right)^{2z}f_0(r) \, dt^2 \, + \, \frac{\ell^2}{r^2 f_0(r)} \, dr^2 \, + \, r^2 \, d\Omega^2_{k,d} \ , \qquad \quad f_0(r) \, = \, 1+\frac{c^2\Phi^2_c}{r^2} \ .
\ee
It loses its dependence on the hyperscaling violating parameter, and also the black hole mass and charge parameters at distance in the order of $(r_h/r)^{d_\theta}$. Again we see how the metric simplifies for large $d$. This time keeping $c^2 \Phi^2$ finite yields $l=\mathcal{O}(1)$. The volume of the horizon $\omega_{k,d}$ goes with $d^{-d/2}$ for spherical and hyperbolical topologies, and $L_\Omega^d$ for planar topology, where $L_\Omega$ is defined as the typical length of the directions $x_0,\ldots,x_{d-1}$. Then taking $G$ small in the large $d$ limit gives no further constraints for spherical and hyperbolical topologies, while it requires $L_\Omega<1$ for planar topology. 

We showed here how the gravitational effects of the black hole decrease outside the horizon. This is due to the phenomenon known in the physics of large dimensional black holes, that is the gravitational potential localizes close to the horizon and the dynamics of the black hole resembles that of a membrane (see e.g.~\cite{Bhattacharyya2015,Bhattacharyya2015a}).

\vspace{5mm}

\section{R\'enyi entropy and quantum corrections}
\label{sec:RenyiQCorr}

In this section we study R\'enyi entropy of Lifshitz scaling spacetimes with quantum corrections perturbatively around $n=1$, and derive constraints based on the null-energy condition and R\'enyi entropy inequalities, which can be used to constrain the characteristics of quantum corrections.

Although Lifshitz scaling spacetimes in $d=2$ exhibit the electric/magnetic duality described in~\S\ref{sec:MagneticElMagnDual}, in the deep IR this duality is expected to break down due to the appearance of corrections to the dilaton potential. Let us first consider the simple solution with zero charge parameter $q=0$ and flat horizon $k=0$. The near horizon geometry of the electric solution is expected to receive $\alpha'$-corrections, while in the case of magnetic solution, quantum corrections become important. This can be seen if we model the quantum corrections as further expansion of the gauge kinetic function and dilaton potential in powers of the coupling $g=\e^{\!\alpha_X \phi}$ such that
\be
X(\phi) \, = \, X_0 \e^{-\!2\alpha_X \phi} \, + \, \xi_1 \, + \, \xi_2 \e^{\!2\alpha_X \phi} \, + \, \ldots \qquad \text{(magnetic solution),} 
\ee
and similarly for $V(\phi)$. Since $\phi$ approaches $-\infty$ near the horizon and $\alpha\leq 0$ provided by the null-energy condition, one can see that these corrections become important for the magnetic solution. Since the electric solution is dually coupled, it does not receive these quantum corrections. The dilaton potential is constant in first non-zero order for both electric and magnetic solution, but possible corrections are taken into account within $V(\phi)$. 

The additional gauge field, which supports the magnetic solution with non-trivial topology ($k\neq 0$) is coupled in the same way as the one above, while the gauge field supporting the non-zero charge parameter ($q\neq 0$) is coupled dually, since $\alpha_Z \geq 0$. Hence one can consider a mixed solution in the sense the first two gauge fields are magnetically charge and the third one is electrically, thus each of them could receive quantum corrections through their gauge kinetic functions.

\subsection{Perturbative solution with quantum corrections} 

In the following derivation we consider arbitrary number of electric and magnetic gauge fields with fluxes $F^{(e)}_i$ and $F^{(m)}_j$, and  general gauge kinetic functions $X_i^{(e)}(\phi)$ and $X_j^{(m)}(\phi)$. We define the functions
\be
W^{(e)}(\phi,r) := \sum_i X_i^{(e)}(\phi) E_i(r)^2  \qquad  \text{and} \qquad W^{(m)}(\phi) := \sum_j X_j^{(m)}(\phi) Q_j^2 
\ee 
for simplicity, where $E_i(r):=(F^{(e)}_i)_{tr}$ is the electric and $Q_j:=(F^{(m)}_j)_{xy}$ is the magnetic field strengths. 

We look for the solution in the form
\be \label{eq:HairyBHansatz}
ds^2 \, = \, L^2 \left(-r^{2z}\e^{2A(r)}f(r) \, dt^2 \, + \, \frac{dr^2}{r^2 f(r)} \, + \, r^2  \, d \Omega^2_{k,2}\right) \ .
\ee
The functions of interest are $A(r)$, $f(r)$ and $\phi(r)$, while $L$ is a scale parameter corresponding to the AdS radius for $z=1$. For Lifshitz spacetime we have $A(r)=z\log r$. The overall spatial scale parameter $\ell$ is chosen to one. We derive the equation of motions by eliminating the second derivative of $f(r)$, and arrive at
\be\begin{aligned}
\mathbf{1.} \quad &\frac{4(r A'(r)-1)}{r^2} \, = \, \phi'{}^{\, 2}(r)   
\ , \\[4mm]
\mathbf{2.} \quad & - 4 L^2 r^2 \Big( r^2 f(r) \big(2 + r A'(r)\big) + r^3 f'(r) - k \Big) \, \\[2mm]
&   \qquad \qquad \qquad \,= \, - \, 2L^4 r^4 V(\phi) \, + \, r^{6} \e^{\!-2A(r)} \, W^{(e)}(\phi,r) \, + \,  W^{(m)} (\phi)    \ , \\[4mm]
\mathbf{3.} \quad & 2 L^2 r^5 \Big( r f(r) \phi''(r) + r f'(r)\phi'(r) + f(r) \phi'(r) (3 + r A'(r)) \Big) \, \\[2mm]
&   \qquad \qquad \qquad \,= \, - \, 2L^4 r^4 \partial_\phi V(\phi) \, - \, r^{6} \e^{\!-2A(r)} \, \partial_\phi W^{(e)}(\phi,r)  \, + \, \partial_\phi W^{(m)} (\phi)   \ .
\end{aligned}\ee
We solve these differential equations perturbatively near the horizon at $r=r_h$, where we impose the boundary conditions
\be
A(r) = A_h + \mathcal{O}(r-r_h) \ , \qquad \quad f(r) = \mathcal{O}(r-r_h) \qquad \text{and} \qquad \phi(r)= \phi_h + \mathcal{O}(r-r_h) \ .
\ee
Perturbing around a Lifshitz solution, which could be done by using the boundary condition $A(r) = z\log r \, ( 1 + \mathcal{O}(r-r_h))$, would not yield an essentially different solution, since they behave similarly near the horizon.
The equations of motion do not give any restriction on $A_h$ as they only depend on the derivatives of $A(r)$.
We use the notation $\varphi(r)=\sum_i \varphi_i (r-r_h)^i$ for the correction terms of an arbitrary function $\varphi(r)$. The first coefficients of the series expansion are
\be \begin{aligned}
\label{eq:PertSolHorizon}
f_1 \, &= \,  \frac{4 k L^2 r_h^2 + 2L^4 r_h^4 V(\phi_h) - r_h^{6} \e^{\!-2A_h} \, W^{(e)}(\phi_h,r_h) - W^{(m)}(\phi_h)}{4 L^2 r_h^5}\ , \\[2mm]
\phi_1 \, & = \, - \, \frac{2}{r_h} \frac{2 L^4 r_h^4 \partial_\phi V(\phi_h) + r_h^{6} \e^{\!-2A_h} \, \partial_{\phi} W^{(e)}(\phi_h,r_h) - \partial_\phi W^{(m)}(\phi_h) }{4 k L^2 r_h^2 + 2L^4 r_h^4 V(\phi_h) - r_h^{6} \e^{\!-2A_h} \, W^{(e)}(\phi_h,r_h) - W^{(m)}(\phi_h)} \ , \\[2mm]
A_1 \, &= \, \frac{1}{r_h} \, + \, \frac{1}{4 r_h} \left(\frac{2 L^4 r_h^4 \partial_\phi V(\phi_h) + r_h^{6} \e^{\!-2A_h} \, \partial_{\phi} W^{(e)}(\phi_h,r_h) - \partial_\phi W^{(m)}(\phi_h)}{4 k L^2 r_h^2 + 2L^4 r_h^4 V(\phi_h) - r_h^{6} \e^{\!-2A_h} \, W^{(e)}(\phi_h,r_h) - W^{(m)}(\phi_h)}\right)^2 \ .
\end{aligned} \ee
The coefficient $A_1$ shows the scaling behavior perturbatively. Since Lifshitz spacetime has $A(r)=z \log r_h + z/r_h (r-r_h)+\mathcal{O}(1-u)^2$, the coefficient $A_1 r_h$ corresponds to the Lifshitz scaling $z$ up to first order, and the quantity defined by $z=1+\Delta z$ shows how far the system is from the relativistic scaling $z=1$, which is
\be
\Delta z \, = \, \frac{1}{4} \left(\frac{2 L^4 r_h^4 \partial_\phi V(\phi_h) + r_h^{6} \e^{\!-2A_h} \, \partial_{\phi} W^{(e)}(\phi_h,r_h) - \partial_\phi W^{(m)}(\phi_h)}{4 k L^2 r_h^2 + 2L^4 r_h^4 V(\phi_h) - r_h^{6} \e^{\!-2A_h} \, W^{(e)}(\phi_h,r_h) - W^{(m)}(\phi_h)}\right)^2   \ .
\ee
One can see that this is zero for AdS spacetime due to vanishing of $\partial_\phi W^{(e)}$, $\partial_\phi W^{(m)}$  and $\partial_\phi V$.

The null-energy condition with null vector
\be
n^\mu \, = \, \left(\frac{1}{\sqrt{f(r)}\e^{A(r)}},r \sqrt{f(r)} \sin\psi,\frac{1}{r}\cos\psi,0\right)
\ee 
results in
\be \begin{aligned}
G_{\mu\nu} n^\mu n^\nu \,  = & \,   \left[ \frac{r_h}{2} (1 + 3 r_h A_1 ) f_1 + r_h^2 f_2 + \frac{k}{ r_h^2}\right]\cos^2\psi \, + \, \mathcal{O}(r-r_h)\cos^2\psi \\
&   \, + \, 2  f_1 (r_h A_1 - 1) (r-r_h)  \, + \, \mathcal{O}(r-r_h)^2 \, \geq \, 0 \ ,
\end{aligned} \ee
which gives two conditions up to first order
\be \label{eq:NECPertExp}
 (r_h A_1 - 1) f_1 \, \geq \,  0 \qquad \qquad \text{and} \qquad \qquad (1 + 3 r_h A_1 ) f_1 + 2 r_h f_2 + \frac{2 k}{r_h^3} \, \geq \, 0  \ .
\ee
The first inequality can be simplified further. The first condition is equivalent to $f_1 \geq 0$ or $r_h A_1 = 1$, which is given in terms of the horizon values as
\be \label{eq:NECPertHor}
 4 k L^2 r_h^2 + 2L^4 r_h^4 V(\phi_h) - r_h^{6} \e^{\!-2A_h} \, W^{(e)}(\phi_h,r_h) - W^{(m)}(\phi_h) \, \geq \, 0 \quad \qquad \text{or} \qquad \quad  z = 1 \ .
\ee
The latter condition means the relativistic scaling. The second inequality gives further constraint on the derivatives $\partial_\phi W^{(e)}(\phi_h,r_h)$, $\partial_\phi W^{(m)}(\phi_h)$  and $\partial_\phi V(\phi_h)$.

\subsection{R\'enyi entropy around $n=1$}

Since the holographic R\'enyi entropy as well as the Bekenstein-Hawking entropy is determined by the horizon geometry, theoretically it is enough to know the solution close to the horizon, however it could be not easy to solve the problem algebraically. In the following we use the perturbative solution derived above to calculate the R\'enyi entropy at first order around $n=1$, and we calculate stability constraint on the quantum corrections of gauge kinetic functions and dilaton potential.

The Hawking temperature of the ansatz \eqref{eq:HairyBHansatz} is expressed as
\be \label{eq:TempGenBHPert}
T \, = \, \frac{1}{4 \pi} \e^{A_h} r_h \, f_1 \, = \, \frac{1}{16 \pi L^2  \, r_h^4} \e^{A_h} \left( 4 k L^2 r_h^2 + 2L^4 r_h^4 V(\phi_h) - r_h^{6} \e^{\!-2A_h} \, W^{(e)}(\phi_h,r_h) - W^{(m)}(\phi_h) \right)  \, ,
\ee
and the Bekenstein--Hawking entropy is
\be
S_{\rm{BH}} \, = \, \frac{\omega_k}{4G} r_h^2 \ .
\ee
Following \eqref{eq:RenyiEntrExpressionInt} we calculate the R\'enyi entropy by
\be
S_n \, = \,  \frac{n}{n-1} \frac{\omega_k}{4G}\frac{1}{T} \int_{T/n}^{T} r_h^2(T') \, d T'  \ .
\ee
The zero order term of the integral $\int_{T/n}^T r_h^2(T) dT$ is non-zero only if the Bekenstein-Hawking entropy is divergent at a given temperature, which would lead to a divergence in the R\'enyi entropy at $n=1$, so we assume that this term is zero. Thus the series expansion at $n=1$ up to first order gives
\be \label{eq:RenyiEntrHorPertFirstOrd}
S_n \, = \, \frac{\omega_k}{4G}\left(r_h^2 \, - \, r_h T \frac{d r_h}{d T}(n-1) \, + \, \mathcal{O}(n-1)^2\right) \ ,
\ee
which is positive, unless $r_h dT/dr_h$ is in the order of $1/(n-1)$.

As an example we calculate the R\'enyi entropy in canonical ensemble perturbatively for the exact Lifshitz scaling solution ($\theta=0$, $d=2$) given in~\S\ref{sec:BackgroundHypScalBH}. The scaling $L^2$ replaces $r_F^{\theta}$ and $\ell$ is chosen to be one, then the R\'enyi entropy up to first order is the same for electric and magnetic solutions
\be
S_n \, = \, \frac{\omega_k r_h^2}{4 G}\left( 1 \, + \, \frac{ q^2 z^2 - z(2 + z) r_h^{2 + 2 z} -k r_h^{2 z}}{q^2 z^2 (2 + z) + z^2 (2 + z) r_h^{2 z+2} + k (z-2) r_h^{2 z} } \, (n-1) \, + \, \mathcal{O}(n-1)^2\right) \ .
\ee

Calculating the R\'enyi inequalities \eqref{eq:RenyiIneq} for \eqref{eq:RenyiEntrHorPertFirstOrd} gives
\be \begin{aligned}
 0 & \ \geq && \frac{\partial S_n}{\partial n}  \, = \, - \frac{\omega_k \, T r_h }{4G} \, \frac{d r_h}{d T} \, + \, \mathcal{O}(n-1) \ ,\\[4mm]
 0 & \ \leq &&  \frac{\partial }{\partial n}\left(\frac{n-1}{n} \, S_n\right)  \, = \, \frac{\omega_k}{4G} \, \frac{r_h^2(T/n)}{n^2} \ , \\[4mm]
 0 & \ \leq && \frac{\partial }{\partial n}\big((n-1) \, S_n\big) \, = \, \frac{\omega_k}{4G} \, r_h^2 \, + \, \mathcal{O}(n-1)\\[4mm]
 0 & \ \geq && \frac{\partial^2 }{\partial n^2}\big((n-1) \, S_n\big) \, = \, - \frac{\omega_k \, T r_h }{4G} \,\frac{d r_h}{d T} \, + \, \mathcal{O}(n-1) \ . \\[4mm]
\end{aligned}\ee
The second and third inequalities, which correspond to the positivity of $S_{\rm{BH}}$, are trivially satisfied, while the other two corresponding to the thermodynamic stability are satisfied up to the first non-zero order if and only if 
\be \label{eq:ineqHorizonTempDer}
\frac{d r_h}{d T} \geq 0 \ ,
\ee
which agrees with the stability condition \eqref{eq:condEntrSpecHeatPos} expanded around $n=1$.

If $d r_h / d T$ is non-zero, the condition \eqref{eq:ineqHorizonTempDer} can be obtained by differentiating \eqref{eq:TempGenBHPert}, which gives
\be \begin{aligned}
\left(\frac{d r_h}{d T}\right)^{-1} \, = \, \frac{\e^{A_h}}{16 \pi L^2\, r_h^5}  & \left[  -  8 k L^2 r_h^2 + 2L^4 r_h^5 \frac{d V(\phi_h)}{d r_h}  -  2 r_h^6 \e^{\!-2 A_h} W^{(e)}(\phi_h,r_h)   \right. \\[2mm]
& \ \  -  r_h^7 \e^{\! -2 A_h} \frac{d W^{(e)}(\phi_h,r_h)}{d r_h}  +  4 W^{(m)}(\phi_h)   - r_h \frac{d W^{(m)}(\phi_h)}{d r_h}  \\[2mm]
& \ \  \left. + \, r_h \frac{d A_h}{d r_h} \left(4 k L^2 r_h^2 + 2L^4 r_h^4 V(\phi_h) - r_h^{6} \e^{\!-2A_h} \, W^{(e)}(\phi_h,r_h) - W^{(m)}(\phi_h)\right) \right] \, .
\end{aligned}\ee
Since the horizon values $A_h$ and $\phi_h$ can both depend on $r_h$, we use the expressions
\be \label{eq:VWderbyScalar}
\begin{aligned}
 \frac{d V(\phi_h)}{d r_h} &= \partial_\phi V(\phi_h) \, \phi_1 \ , \qquad \qquad  \frac{d W(\phi_h)}{d r_h} = \partial_\phi W(\phi_h) \, \phi_1  \ , \qquad\qquad  \frac{d A_h}{d \phi_h} = A_1 \ ,   \\[2mm]
\frac{\partial W(\phi_h,r_h)}{\partial r_h} &=  2 \left(A_1 - \frac{3}{r_h}\right) W(\phi_h,r_h) - 2 \phi_1 \partial_\phi W(\phi_h,r_h) \ .
\end{aligned}\ee
To derive the last one we used the expansion of the Maxwell equation
\be
E_i'(r_h) \, = \, \left( A_1 - \frac{3}{r_h} -  \frac{\partial_\phi X_i(\phi_h) }{X_i(\phi_h) }  \phi_1 \right)  E_i(r_h) \ .
\ee
Then the stability constraint \eqref{eq:ineqHorizonTempDer} together with the first null-energy condition of \eqref{eq:NECPertExp} can be rewritten in the final form
\be \label{eq:GenPotIneqRenyi}
\begin{aligned}
0 \, & \leq \, \frac{\left(2L^4 r_h^4 \partial_\phi V(\phi_h) + r_h^6 \e^{\!-2 A_h} \, \partial_\phi W^{(e)}(\phi_h,r_h)  - \partial_\phi W^{(m)}(\phi_h) \right)^2}{ 4 k L^2 r_h^2 + 2L^4 r_h^4 V(\phi_h) - r_h^{6} \e^{\!-2A_h} \, W^{(e)}(\phi_h,r_h) - W^{(m)}(\phi_h)} \, \\[2mm]
& \qquad \quad \leq \, - 4 k L^2 r_h^2 + 2L^4 r_h^4 V(\phi_h) + 3 r_h^{6} \e^{\!-2A_h} \, W^{(e)}(\phi_h,r_h) + 3 W^{(m)}(\phi_h) \ .
\end{aligned}\ee

The general formula above can give constraint on finite quantum corrections to the Lifshitz and Hyperscaling violating solution studied in~\S\ref{sec:BackgroundHypScalBH}. These corrections can considered as extra terms in $W^{(e)}(\phi_h,r_h)$, $W^{(m)}(\phi_h)$ and $V(\phi_h)$ as
\be \begin{aligned}
W^{(e)}(\phi_h,r_h) \, &= \, \sum_i X_{0,i}^{(e)} \e^{\!2\alpha^{(e)}_i \phi_h} \, E_i(r_h)^2 \, + \, W^{(e)}_{1}(\phi_h,r_h) \ , \\[2mm]
W^{(m)}(\phi_h) \, &= \, \sum_j X_{0,j}^{(m)} \e^{-\!2\alpha^{(m)}_i \phi_h} \, Q_j^2 \, + \, W^{(m)}_{1}(\phi_h) \ , \\[2mm]
V(\phi_h) \, &= \, V_0 \e^{\! \eta \phi_h} + V_{1}(\phi_h) \ .
\end{aligned}\ee
The signs of $\alpha^{(e)}_i$ and $\alpha^{(m)}_j$ determine whether the IR dynamics is captured by quantum correction in the gauge kinetic function or not. Let us assume $\alpha^{(e)}_i > 0$, while $\alpha^{(m)}_j < 0$, so all of them are expected to receive quantum corrections. If we take them into account as exponential expansions
\be \begin{aligned}
W^{(e)}_{1}(\phi_h,r_h) \, &= \, \sum_i \left( \xi_{1,i}^{(e)} \, + \, \xi_{2,i}^{(e)} \e^{-\!2\alpha^{(e)}_i \phi_h} \, + \, \ldots \right) E_i(r_h)^2 \ , \\[2mm]
W^{(m)}_{1}(\phi_h) \, &= \, \sum_j \left( \xi_{1,i}^{(m)} \, + \, \xi_{2,j}^{(m)} \e^{\!2\alpha^{(m)}_i \phi_h} \, + \, \ldots  \right) Q_j^2  \ , \\[2mm]
V_{1}(\phi_h) \, &= \, \rho_1 \, +  \, \rho_2 \e^{\! - \eta \phi_h}  \, + \, \ldots  \ ,
\end{aligned} \ee 
then the formula \eqref{eq:GenPotIneqRenyi} constrains the finite correction coefficients $\xi^{(e)}_{k,i}$, $\xi^{(m)}_{k,j}$ and $\rho_k$.

\vspace{5mm}

\section{Conclusion and outlook}

In this paper we studied holographic R\'enyi entropy of Lifshitz and hyperscaling violating black hole solutions in Einstein-Maxwell-dilaton gravity. In~\S\ref{sec:CalcRenyiEntr} we analyzed the R\'enyi entropy inequalities for different values of Lifshitz parameter, horizon topology and electric potential, which in some cases led to upper or lower bound for the parameters. We saw the inequalities have a close connection to thermodynamic stability of the black hole, but they do not tell us about the Hawking-Page phase transition for spherical horizon topologies. When the Lifshitz parameter is $1\leq z < 2$ and the electric potential is smaller than $\Phi_c$ defined in \eqref{eq:PhicDef}, the maximal value \eqref{eq:nmaxBH} of R\'enyi parameter is larger than the parameter corresponding to the critical horizon value in Hawking-Page phase transition. Hence the effect of the phase transition can occur within the possible parameter values. Since this phenomena is not captured by the holographic R\'enyi entropy we studied in this paper, it would be interesting to investigate the dual R\'enyi entropy in this context.

We calculated R\'enyi entropy for specific values of the Lifshitz scaling parameter $z$, and analyzed the dual ground state degeneracy and value by studying the $n=1$ and $n=\infty$ limits of R\'enyi entropy. A further study could compare our results with QFT calculations (see e.g.~\cite{Mollabashi2017,He2017} for entanglement measures in Lifshitz scaling scalar field theories). An interesting further direction would be a more detailed study of the dual spectrum by reconstructing it from R\'enyi entropies. Although it may be a difficult problem to work	out analytically, it could be studied numerically (e.g.~by Laplace transformation or the theory of symmetric polynomials for discrete spectrum). 

By analyzing the R\'enyi entropy we found that the dual spectrum simplifies in the limit when $d_\theta=d-\theta$ approaches infinity. The degeneracy of the ground state goes to one if the square of the electric potential $\Phi^2$ is smaller than $\Phi_c^2+\ell^2/c^2$. We also gave some remark on the background geometry in the parameter limits, which can realize large $d_\theta$. We pointed out an interesting phenomenon, which is known in the context of large dimensional black holes in general relativity or AdS spacetime. They were studied in detail over the past few years (see e.g.~\cite{Emparan2013,Emparan2015,Bhattacharyya2015,Emparan2015a,Bhattacharyya2015a}), and it was found that in this limit the gravitational field of a black hole is strongly localized near its horizon, and the black hole can be replaced by a membrane. As a further direction, the large $d_\theta$ limit could be studied more thoroughly in this context. Another double scaling limit was suggested in \cite{Hartnoll2012}, in which both $\theta$ and $z$ approach infinity, with their ratio held fixed. In this limit the entropy behaves nicely in the sense that it vanishes for ground state. It would be interesting to combine these limits in further study. 

In~\S\ref{sec:RenyiQCorr} we solved the Einstein-Maxwell-dilaton equations of motions with general gauge kinetic functions and dilaton potential perturbatively, which was motivated by the goal of including quantum corrections. We calculated the R\'enyi entropy of Lifshitz scaling solutions in canonical ensemble around $n=1$, and derived constraints on the gauge kinetic functions and dilaton potentials using the R\'enyi entropy inequalities, which correspond to the thermodynamic stability of the black hole. Our general result could be used to specify the range of quantum corrections in more detail. Another possible direction would be to include $\alpha'$ corrections. In this paper we studied the R\'enyi entropy perturbatively around $n=1$ up to first order, and the two non-zero terms corresponded to entropy and heat capacity respectively. It could be interesting to study the higher order terms and their relation to thermodynamic quantities.

\section*{Acknowledgments}

This work was supported by the Hungarian research grant NKFIH K116505. The work of Z.K. was supported by the Croatian Science Foundation Project ``New Geometries for Gravity and Spacetime'' (IP-2018-01-7615), and also partially supported by the European Union through the European Regional Development Fund - The Competitiveness and Cohesion Operational Programme (KK.01.1.1.06).

\appendix
\section{An example for both discrete and continuous spectrum}
\label{sec:ExDiscContSpec}

We calculate the energy density of the dual modular Hamiltonian in the classical holographic limit with parameters $z=1$, $\theta=d-1$ and $\omega_{k,d}=4G$, $r_F=\ell=1$ for simplicity. The R\'enyi entropy does not depend on the electric potentials. Then the partition function with R\'enyi paramter can be written as
\be
Z_n = \int_0^\infty d E \, \rho(E) \e^{\! -nE/T} \ ,
\ee 
where $\rho(E)$ is the energy density of eigenvalues $E$ of the corresponding modular Hamiltonian, and $Z_1=\e^{\! - G/T}$, where $G$ is the thermodynamic potential. The energy density have non-zero values from a minimal energy $E_{\rm{min}}$. Then $Z_n$ can be expressed with the R\'enyi entropy as follows
\be
Z_n = \e^{\! (1-n)S_n -n G/T} = \e^{\! - \pi T \, \frac{(n-1)(n+1)}{n} - \frac{nG}{T}} \ .
\ee
One can calculate the energy density $\rho(E)$ as an inverse Laplace transformation, which gives
\be
\rho(E) \, = \, \Theta\Big(\frac{E-G-\pi T^2}{T}\Big)\Bigg[\frac{\sqrt{\pi}}{2}\frac{I_1\left(\sqrt{\pi\left(E-G-\pi T^2\right)}\right)}{\sqrt{E-G-\pi T^2}} + \frac{1}{T}\,\delta\bigg(\frac{E-G-\pi T^2}{T}\bigg)\Bigg] \ ,
\ee
where $\Theta(z)$ is the Heaviside function, $I_1(z)$ is the modified or hyperbolic Bessel function of first kind and $\delta(z)$ is the Dirac delta. Then one can rewrite $Z_n$ as
\be
Z_n = \int_0^\infty d E \, \rho^*(E) \e^{\! -n(E+G+\pi T^2)/T}  \qquad \text{with} \qquad \rho^*(E) =  \frac{\sqrt{\pi}\, I_1\left(2\sqrt{\pi E}\right)}{\sqrt{E}} + \frac{1}{T}\, \delta\left(\frac{E}{T}\right)  \ ,
\ee
which indicates a ground state with energy $E_{\rm{gr}}=G+\pi T^2$, which could be expected by calculating the limit $S_\infty=(E_{\rm{gr}}-G)/T$. Its degeneracies are given by both discrete $\delta(E/T)$ and continuous $\lim_{E\rightarrow 0}\sqrt{\pi} I_1(2\sqrt{\pi E})/\sqrt{E}= \pi$ spectral densities.


\end{document}